\documentclass[twocolumn]{aastex631}

\usepackage{subfigure}
\usepackage{amsmath}
\usepackage{esint}

\shorttitle{CX/OX Structures in Boxy-Peanut Bulges}
\shortauthors{B. Tahmasebzadeh et al.}
\graphicspath{{./}{figures/}}
\begin{document}

\title{Orbital Support and Evolution of CX/OX Structures in Boxy/Peanut Bars}

\author{Behzad Tahmasebzadeh}\thanks{behzad@umich.edu,}
\affiliation{Department of Astronomy, University of Michigan, Ann Arbor, MI, 48109, USA}
%\affiliation{Shanghai Astronomical Observatory, Chinese Academy of Sciences, 80 Nandan Road, Shanghai 200030, China}

\author{Shashank Dattathri}
\affiliation{Department of Astronomy, Yale University, Kline Tower 266 Whitney Avenue, New Haven, CT 06511, USA}

\author{Monica Valluri}
\affiliation{Department of Astronomy, University of Michigan, Ann Arbor, MI, 48109, USA}

\author{Juntai Shen}%\thanks{Corr author: jtshen@sjtu.edu.cn;} 
\affiliation{Department of Astronomy, School of Physics and Astronomy, Shanghai Jiao Tong University, 800 Dongchuan Road, Shanghai 200240, China}
\affiliation{Key Laboratory for Particle Astrophysics and Cosmology (MOE) / Shanghai Key Laboratory for Particle Physics and Cosmology, Shanghai 200240, China}

\author{Ling Zhu} %\thanks{lzhu@shao.ac.cn.} 
\affiliation{Shanghai Astronomical Observatory, Chinese Academy of Sciences, 80 Nandan Road, Shanghai 200030, China}

\author{Vance Wheeler}
\affiliation{Department of Physics, University of Chicago, Chicago, Illinois 60637, USA}

\author{Ortwin Gerhard}
\affiliation{Max-Planck-Institut f{\"u}r Extraterrestrische Physik, Gie{\ss}enbachstra{\ss}e 1, 85748 Garching, Germany}

%\author[0000-0001-7902-0116]{Victor Debattista}
%\affiliation{Jeremiah Horrocks Institute, University of Central Lancashire, Preston PR1 2HE, UK}

\author{Sandeep Kumar Kataria}
\affiliation{Department of Astronomy, School of Physics and Astronomy, Shanghai Jiao Tong University, 800 Dongchuan Road, Shanghai 200240, China}

\author{Leandro {Beraldo e Silva}}
\affiliation{Department of Astronomy \& Steward Observatory, University of Arizona, Tucson, AZ 85721, USA}

\author{Kathryne J. Daniel}
\affiliation{Department of Astronomy \& Steward Observatory, University of Arizona, Tucson, AZ 85721, USA}
\affiliation{Center for Computational Astrophysics, Flatiron Institute, New York, NY 10010, USA}
%% Note that the \and command from previous versions of AASTeX is now
%% depreciated in this version as it is no longer necessary. AASTeX 
%% automatically takes care of all commas and "and"s between authors names.

%% AASTeX 6.31 has the new \collaboration and \nocollaboration commands to
%% provide the collaboration status of a group of authors. These commands 
%% can be used either before or after the list of corresponding authors. The
%% argument for \collaboration is the collaboration identifier. Authors are
%% encouraged to surround collaboration identifiers with ()s. The 
%% \nocollaboration command takes no argument and exists to indicate that
%% the nearby authors are not part of surrounding collaborations.

%% Mark off the abstract in the ``abstract'' environment. 
\begin{abstract}
Barred galaxies exhibit boxy/peanut or X-shapes (BP/X) protruding from their disks in edge-on views. Two types of BP/X morphologies exist depending on whether the X-wings meet at the center (CX) or are off-centered (OX). Orbital studies indicate that various orbital types can generate X-shaped structures. Here, we provide a classification approach that identifies the specific orbit families responsible for generating OX and CX-shaped structures. Applying this approach to three different N-body bar models, we show that both OX and CX structures are associated with the x1 orbit family, but OX-supporting orbits possess higher angular momentum (closer to x1 orbits) than orbits in CX structures. Consequently, as the bar slows down the contribution of higher angular momentum OX-supporting orbits decreases and that of lower angular momentum orbits increases resulting in an evolution of the morphology from OX to CX. If the bar does not slow down, the shape of the BP/X structure and the fractions of OX/CX supporting orbits remain substantially unchanged. Bars that do not undergo buckling but that do slow down initially show the OX structure and are dominated by high angular momentum orbits, transitioning to a CX morphology. Bars that buckle exhibit a combination of both OX and CX supporting orbits immediately after the buckling, but become more CX dominated as their pattern speed decreases. This study demonstrates that the evolution of BP/X morphology and orbit populations strongly depends on the evolution of the bar angular momentum.

\end{abstract}


%-----------------------------------------------------
\section{Introduction}

In numerous observations of external disk galaxies viewed edge-on, boxy/peanut or X-shaped bulges (hereafter referred to as BP/X bulges) have been identified \cite[]{ Shaw.1987, Ltticke.2000, Erwin.2017, Li.2017}. Within the Milky Way, a distinct BP/X-shaped structure was first discerned in the multi-parameter model of COBE/DIRBE images of the Galactic Bulge \cite[]{Freudenreich.1998}. 
\par

The BP/X bulges have been consistently observed in a variety of N-body simulations. %suggesting a correlation with the vertical growth of bars. 
Early simulations showed that BP/X bulges can form following a short-lived buckling event in a bar which results from an asymmetric bending of the bar out of the disk midplane  \cite[]{Combes.1990,Pfenniger.1991,Raha.1991,Merritt.1994}.  
However, recent evidence suggests that orbital resonances may play a crucial role in forming and enhancing BP/X structures. One scenario for the formation of BP/X structures without a buckling event is 'resonant trapping,' where orbits are vertically excited by being trapped at the vertical Inner Lindblad Resonance (vILR) for significant periods \citep{Quillen.2002}. Another mechanism is 'resonant sweeping,' in which orbits cross the vILR, become vertically heated, and remain that way after leaving the resonance \citep{Quillen.2014}.  \cite{Sellwood.2020} reviewed all three mechanisms for the formation of BP/X structures and found evidence for the resonant trapping mechanism only in an artificially vertically symmetrized model. They concluded that the resonant sweeping mechanism is likely more applicable in real galaxies without bar buckling. This conclusion was later confirmed by \cite{Wheeler.2023} and \cite{Beraldo.2023}.

\par
\begin{figure*}
	\centering	%
	\includegraphics[width=2\columnwidth]{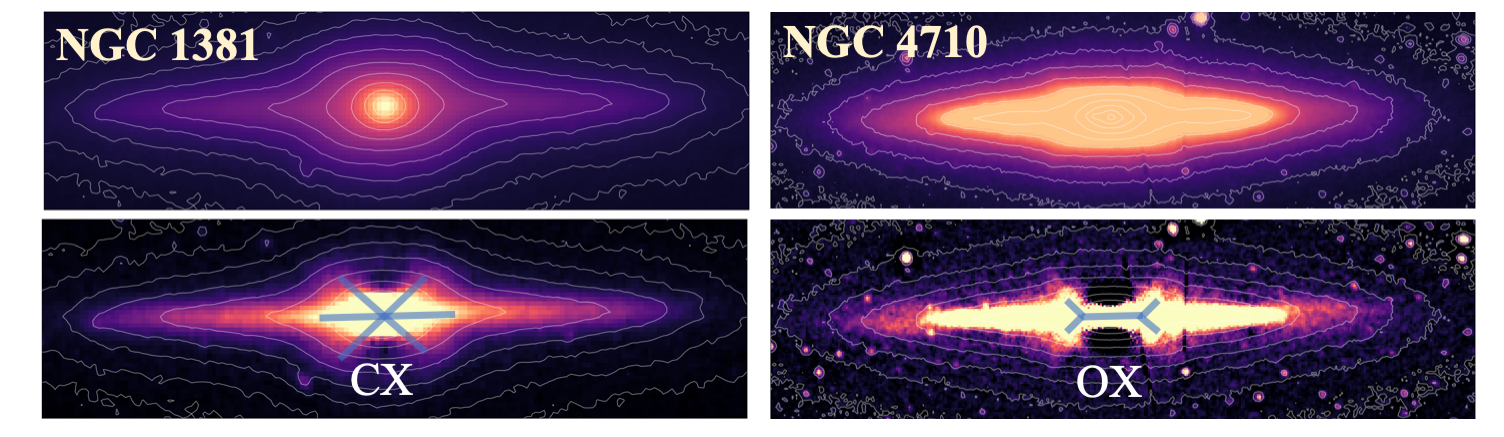}
	\hspace{8pt}%
	\caption{The top row shows the S$^4$G images for NGC 1381 and NGC 4710. The bottom row shows the unsharp masked version of each image. The BP/X-shaped structure in NGC 1381 is predominantly of the CX shape, while NGC 4710 displays an OX structure. }%
	\label{fig:s4g}%
\end{figure*}

BP/X bulges are classified into two categories based on their morphology, as described by \cite{Bureau.2006}. The two categories are the off-centered X-shape (OX), where the X-wings do not intersect at the center (which look like $>-<$), and the centered X-shape (CX), where the X-wings cross at the center and in the disk plane ($><$). Figure \ref{fig:s4g} presents examples of an OX bulge (NGC 1381) and a CX bulge (NGC 4710), displaying their S$^4$G images with the corresponding Gaussian-filtered unsharp masked versions.
\par
Orbital analysis is a crucial tool for deciphering the building blocks of barred galaxies. While the orbital composition of BP/X bulges has been extensively studied, several aspects remain contentious. Key among these is the identification of specific orbit families that independently support CX and OX structures. Another is understanding how the evolution of the BP/X structure correlates with changes in the bar's characteristics, such as its pattern speed.

\par

In many studies of three-dimensional analytical bar models \cite[]{Pfenniger.1991, Skokos.2002, Patsis.2002, Patsis.2018}, the periodic orbits bifurcating from the $x_{1}$ family (referred to as $x_{1}$-tree orbits) are considered the backbone of X-structures. This includes orbit families such as $x_{1}v_{1}$ ($\smile$ or $\frown$) and $x_{1}v_{2}$ ($\infty$), which are associated with the $(\Omega_{z}:\Omega_{x} = 2:1)$ resonance. 

\par

However, some other studies questioned these conclusions. \cite {Portail.2015} analyzed the orbital structure of Made-to-Measure (M2M) models for the Milky Way bar. They found 3D resonant orbits with $(\Omega_{z}:\Omega_{x} = 2:1)$ or higher vertical resonances of the $ x_{1} $ orbit family cannot fully explain the X-shape of the bar. In their models, the fraction of $(\Omega_{z}:\Omega_{x} = 2:1)$ resonant orbits is relatively small, and they are predominantly located in the outer regions of the bar. Consequently, they proposed an alternative family of resonant boxlet orbits, termed `brezel orbits' associated with the $(\Omega_{z}:\Omega_{x} = 5:3)$ resonance. These brezel orbits are posited to generate the X-shape in the inner regions of the bar. 
\par
%The composition of different orbit families in N-body bars is studied by \cite{Valluri.2016}. The most dominant orbits family in their models is box family ($ \sim 60 \% $) rather than $x_{1}$ family. 

\cite{Abbott.2017} studied an N-body bar model to explore the orbits responsible for building the BP/X bulge. They also found a few $ x_{1}v_{1} $ orbits ($ \sim 3 \% $) which are in the outer half of the bar (similar to \cite {Portail.2015} and \cite{Valluri.2016}). The most populated resonant boxlet family in their N-body bar is fish/pretzel orbits ($ \sim 6 \% $) that are associated with the $(\Omega_{x}:\Omega_{y}:\Omega_{z} = 3:-2:0)$ resonance. They argued that no individual orbit family is the backbone of the BP/X bulge. They found that non-resonant box orbits  ($ \sim 63 \% $), banana orbits  ($ \sim 3 \% $), fish/pretzel orbits  ($ \sim 6 \% $), and brezel orbits ($ \sim 1.5 \% $), each contributes to the formation of X-shaped structures.  This finding underscores the relatively minor role of resonant orbits in making up the BP/X structure.
\par
\cite{Parul.2020} confirmed that various types of orbits can support the BP/X structures. They examined the orbital structures of two N-body bar models. They classified the bar orbits only based on the ratio of the vertical oscillation frequency to the in-plane frequency ($\Omega_{z} / \Omega_{x}$). Three groups of orbits are considered: %$\Omega_{z} / \Omega_{x} = 1.55-1.75$, $\Omega_{z} / \Omega_{x} = 1.75-1.95$, and  $\Omega_{z} / \Omega_{x} = 1.95-2.05$.
$1.55<\Omega_{z} / \Omega_{x} \leq 1.75$, $1.75<\Omega_{z} / \Omega_{x} \leq 1.95$, and  $ 1.95<\Omega_{z} / \Omega_{x} \leq 2.05$. Their findings demonstrated that each group is capable of forming an X-shaped structure, as illustrated in Fig. 9 in \cite{Parul.2020}. 
They concluded that X-shaped structures are not formed by specific orbits serving as the backbone. Instead, these structures arise from the assembling of high-density regions of different types of orbits at their highest points. 
\par
The orbital composition of the CX and OX structures is discussed only in a few studies. \cite{Patsis.2018} showed that periodic orbits around $ x_{1}v_{2} $ support a CX profile (see Fig. 6 in \cite{Patsis.2018}), while the sticky-chaotic orbits with initial conditions close to $ x_{1}v_{1} $ and 3D quasi-periodic orbits around $ x_{1} $ support an OX structure (see Fig. 7 in \cite{Patsis.2018}). They also analyzed an N-body bar model presented in \cite{Contopoulos.2013}, and found that sticky chaotic orbits support an OX profile, and periodic orbits build a CX structure. \cite{Parul.2020} showed the orbits with higher $\Omega_{z} / \Omega_{x}$ are generating the OX shape, and those with lower $\Omega_{z} / \Omega_{x}$ are contributing to the CX structure (see Fig. 9 in \cite{Parul.2020}). \cite{Valencia.2023} conducted an orbital frequency analysis of live N-body models and confirmed that orbits with varying ranges of $\Omega_{z} / \Omega_{x}$ contribute to the formation of an X-shaped structure. They found that during the initial stages of bar formation, prior to the buckling phase, the ratios $\Omega_{z} / \Omega_{x}$ are higher. As the models evolve, the distribution of $\Omega_{z} / \Omega_{x}$ shifts towards lower values. See Also \cite{Sellwood.2020}.

\begin{figure}
	\centering	%
	\includegraphics[ width=\columnwidth]{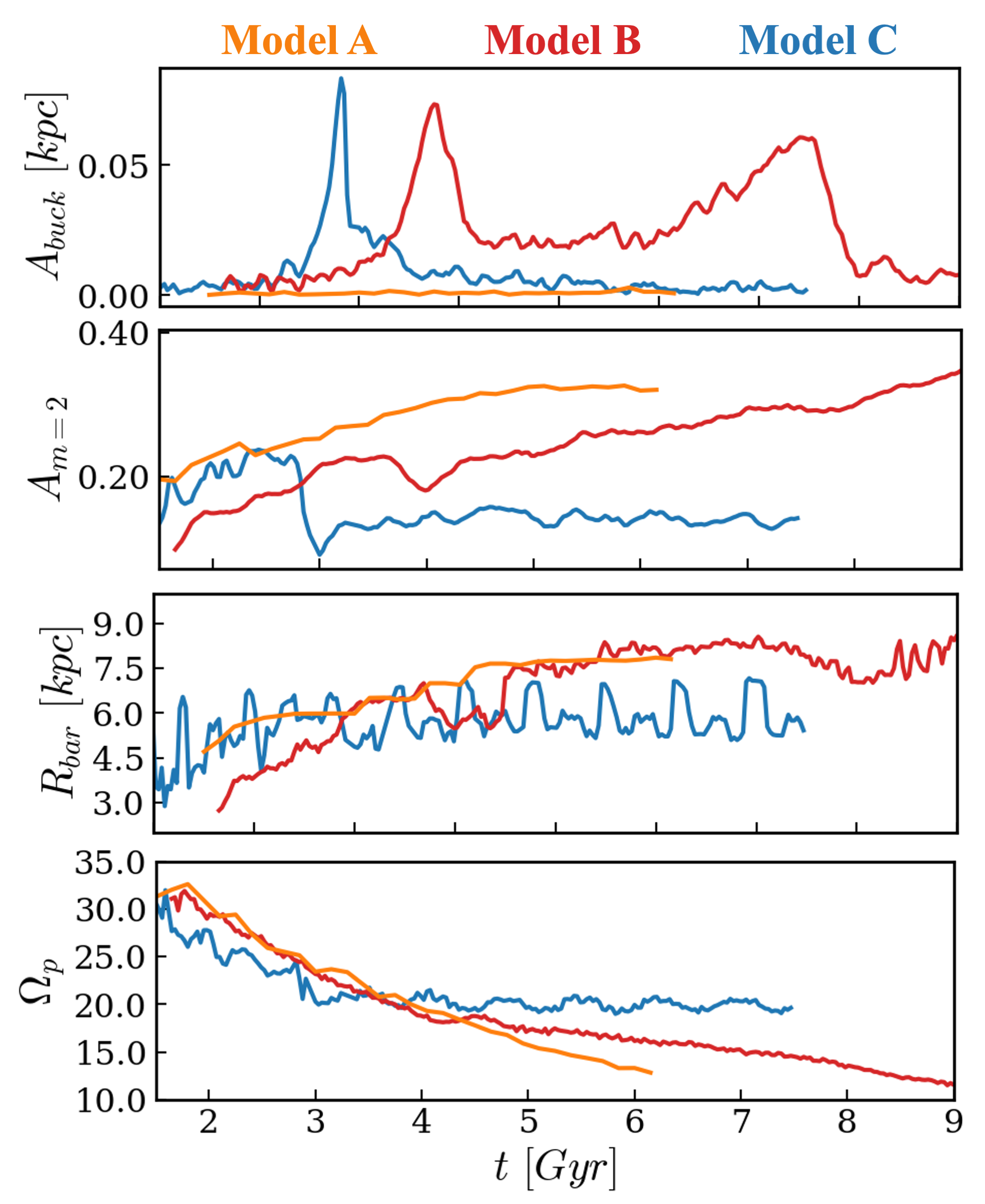}
	\caption{ The globally measured bar quantities as a function of time for model A (orange), model B (red), and model C (blue). The panels from top to bottom represent: (1) the buckling amplitude ($A_{\rm buck}$), which is the scaled $m=2$ asymmetry about the midplane; (2) the bar amplitude ($A_{\rm m}=2$), which is the scaled $m=2$ Fourier amplitude; (3) the bar length; and (4) the bar pattern speed. For model C, the $A_{\rm m}$ values are multiplied by a factor of 0.75, and the pattern speed is multiplied by a factor of 0.3 for better visualization. }%
	\label{fig:model_param}%
\end{figure}

\begin{figure*}
	\centering	%
	\includegraphics[width=2\columnwidth]{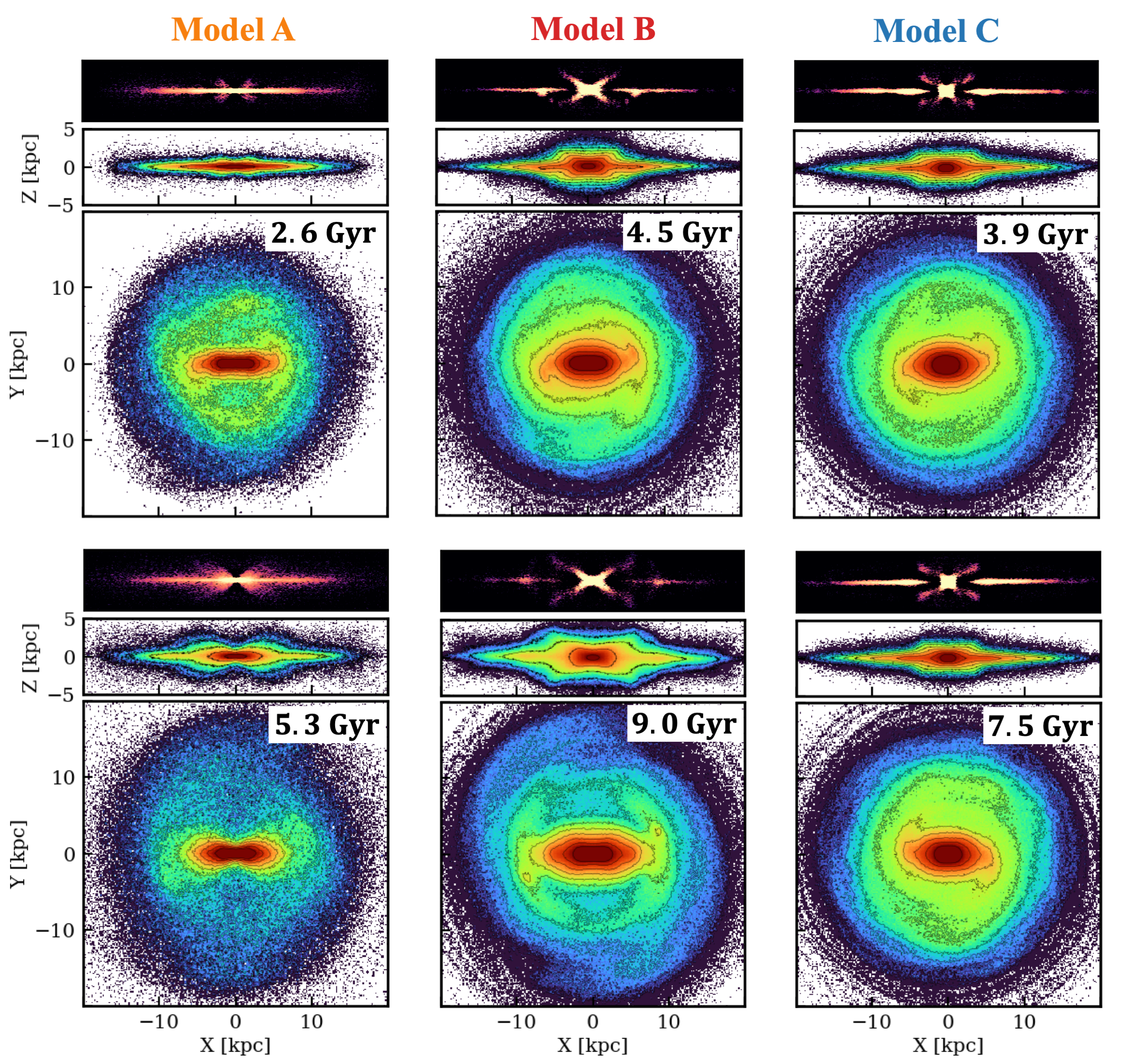}
	\caption{Unsharp masked image and projected surface density in face-on and edge-on views for model A (left), model B (middle), and model C (right). The top row shows snapshots taken shortly after the formation of the BP/X structure, while the bottom row represents snapshots taken around the end of the simulation. }%
	\label{fig:model}%
\end{figure*}

\par
In this work, we revisit the orbital composition in BP/X bulges. While it is now understood that various orbit types can give rise to X-shaped structures, yet a universal classification applicable across models with diverse characteristics has not been established. Moreover, there is no clear explanation of what determines the differences between OX and CX structures, nor an understanding of how their orbital compositions differ.

Here, we employ an automated orbit classification method based on frequency analysis \cite[]{Valluri.1998, Valluri.2016} to systematically investigate the orbit families that give rise to the OX and CX-shaped structures. Additionally, we explore the influence of the bar pattern speed on the development of OX/CX structures, as well as the types of orbits that constitute bars with different pattern speeds.

\par
The paper is organized as follows. In Section \ref{Nbody}, we provide
the details of the simulations used in this study. Section \ref{method} presents the description of the orbital analysis and classification method. The orbital structure of the BP/X bulges is studied in Section \ref{Xshape}, and the discussion follows in Section \ref{conl}.

\section{N-body Bar Models} \label{Nbody}
We study three N-body Milky Way-like models evolved using the GALAXY code \citep{Sellwood.2014}. Model A, was kindly provided by J. Sellwood and is referenced as model C in \cite{Sellwood.2020}. This model consists of $10^{6}$ equal-mass particles representing an exponential disk with a total mass of $4.21 \times 10^{10} M_{\odot}$, and $10^{6}$ particles as a live isotropic spherical Hernquist halo with a total mass of $4.21 \times 10^{11} M_{\odot}$. This model is evolved for 6 Gyr, and the buckling instability is inhibited by imposing reflection symmetry about the mid-plane at each step of the simulation \citep[see][for details]{Sellwood.2020}.

We also study model run6000 and model C from \cite{Wheeler.2023}, which we refer to as model B and model C, respectively. Both models evolved under the same simulation setup. The first 9 Gyr of the evolution of Model B is used, after which it soon reaches a steady state, and model C evolved for 7.5 Gyr. See \cite{Wheeler.2023} for details. 

Both model B and model C comprised $6 \times 10^{6}$ particles representing an exponential disk with a total mass of $5.37 \times 10^{10} M_{\odot}$, and $4 \times 10^{6}$ particles in a live Navarro–Frenk–White (NFW) halo \cite[]{NFW.1996} with a total mass of $6.77 \times 10^{11} M_{\odot}$. Model B is characterized by a higher initial radial velocity dispersion compared to model C.

To quantify and compare the properties of the bars, we utilize standard measures, bar amplitude ($A_m = 2$) and buckling amplitude ($A_{\rm buck}$), following e.g., \cite{Debattista.2020}. These parameters are defined by employing the $m = 2$ symmetry mode in an azimuthal Fourier expansion of the disk viewed face-on. These quantities are  normalized by the $m = 0$ mode and defined as follows:

\begin{equation}
A_{\mathrm{m}=2}=\left|\frac{\sum_k m_k e^{2 i \phi_k}}{\sum_k m_k}\right|
\end{equation}

and 

\begin{equation}
A_{\text {buck }}=\left|\frac{\sum_k z_k m_k e^{2 i \phi_k}}{\sum_k m_k}\right|.
\end{equation}
where $m_k, \phi_k$, and $z_k$ are the mass, azimuth, and vertical position of the $k$ th particle, respectively. We measure the pattern speed of the models over time using the approach developed by \cite{Dehnen.2023}, which makes it feasible to determine the pattern speed using a single snapshot.  Fig. \ref{fig:model_param} presents the evolution of several bar properties: buckling amplitude ($A_{\text{buck}}$), bar amplitude ($A_{\mathrm{m}=2}$),  bar length, and bar pattern speed ($\Omega_p$), for model A (orange), model B (red), and model C (blue).

Model A does not undergo a buckling event, and the BP/X structure gradually forms, becoming more pronounced by the end of the simulation. The bar strengthens and grows until $\sim 5$ Gyr, and continuously slows down by the end of the simulation. Model B experiences two buckling events, approximately at 3.8 Gyr, and 7.7 Gyr, followed by a gradual increase in bar amplitude and associated decrease in pattern speed throughout the evolution. In contrast, model C undergoes only one strong buckling event early in its evolution at around 2.8 Gyr after which the bar ceases to strengthen and grow and maintains a steady pattern speed until the end of the simulation.

Figure \ref{fig:model} illustrates the face-on and edge-on projected surface densities alongside the unsharp masked images for model A (left), model B (middle), and model C (right). The top row displays snapshots taken soon after the formation of the BP/X structure, while the bottom row presents snapshots from around the end of the simulation. At first glance, it is evident that the BP/X shape has undergone significant evolution in model A and model B, but shows minimal alteration in model C.  

\par

\section{ORBIT ANALYSIS METHODS}
\label{method}
\subsection{Orbit integration}
In this section, we examine the orbital structure of all the snapshots shown in Fig. \ref{fig:model}. Detailed analysis plots will be provided exclusively for model B at $t=4.5$ Gyr. For the remaining models, only the final results will be presented to avoid an excess of figures.

We employ AGAMA \footnote{\url{https://github.com/GalacticDynamics-Oxford/Agama}} \cite[]{Vasiliev.2019} to compute the potential and orbits in our N-body models. The N-body system is frozen at the specified snapshots, and then potentials are calculated from the particle distribution using \texttt{CylSpline} expansion \cite[]{Binny.2008}. 
\par

We randomly selected 15,000 initial conditions corresponding to the positions and velocities of particles from the snapshots. These orbits are then integrated over a duration of $20 \hspace{.05cm} \mathrm{Gyr}$ ($\sim$ 200 orbital period at the end of the bar region) in the presence of the rotating bar, with the given $\Omega_{\mathrm{p}}$ for each snapshot. We store 10,000 points per orbit with equal time intervals. 
We choose a long integration time to better visualize the different bar structures discussed in the following section and to ensure that orbits in the outer regions are integrated for a sufficient period to compute frequencies accurately. However, we also test our results by performing the integration over a much shorter time of 2 Gyr ($\sim$  20 orbital periods at the end of the bar region) and demonstrate that the integration time does not significantly affect our conclusion (see Appendix \ref{Appendix}).

%The time period of a circular orbit at the end of the bar region is around $ \mathrm{t} = 0.1 \hspace{.05cm} \mathrm{Gyr} $ in model A, and around $ \mathrm{t} = 0.8 \hspace{.05cm} \mathrm{Gyr} $ for model B.

\subsection{Orbits classification}
Frequency analysis of orbits plays a crucial role in deciphering the characteristics of orbital structures within a large sample. To compute the fundamental frequencies in Cartesian and cylindrical coordinates, we utilize the NAFF (Numerical Analysis of Fundamental Frequencies) software \footnote{\url{https://bitbucket.org/cjantonelli/naffrepo/src/master/}}  \cite[]{Valluri.1998, Valluri.2016}.
\par
Automated classification of orbits, based on fundamental frequencies, is a well-developed technique that has been widely utilized in the literature \cite[]{Carpintero.1998,Valluri2.2010,Vasiliev.2013,Valluri.2016}. As discussed in \cite{PatsisAna.2019}, orbit classification based solely on frequency ratios is insufficient for a comprehensive orbital structure framework. This is because orbit families with identical frequency ratios can exhibit divergent properties, including variations in energy ranges, stability, extent, and shapes. In contrast, the auto-classification method in NAFF incorporates multiple quantities in addition to the orbital frequencies, such as the apocenter radius of orbits, the maximum values of the $x$ and $y$ coordinates, and the net rotation parameter. See appendix B in \cite{Valluri.2016} for details. A new implementation of NAFF code named naif \footnote{\url{https://naif.readthedocs.io/en/latest/index.html}} has been publicly released in a python package with some new features by \cite{Beraldo.2023}. The orbit auto-classifier and resonance auto-finder for frequency maps are currently being developed for inclusion in the naif package (R. Ranjan et al., 2024, in preparation).
\par
To determine whether an orbit is prograde or retrograde in the bar's reference frame, we use the following expression, which we call the net rotation parameter:

\begin{equation}
\overline{L}_{zn} =  \frac{ \Sigma_{i=1}^{i=N} \hspace{0.1cm} \mathrm{sign}(L_{z})_{i} }{N},  
 \hspace{1cm}
\end{equation}	
where N is the number of time steps. For a positive and negative value of $L_{z}$, $\mathrm{sign}(L_{z})$ is $+1$ and $-1$, respectively. A short-axis orbit with $\overline{L}_{zn}\sim 1.0$ is prograde with the highest angular momentum, while $\overline{L}_{zn}\sim -1.0$ indicates a retrograde orbit around the z-axis in the bar rotating frame. An orbit with no net rotation around the z-axis has $\overline{L}_{zn}\sim 0.0$. A long-axis tube orbit has $\overline{L}_{xn}\sim \mp 1.0$ since it has net rotation around the x-axis.
\par 
In this approach, disk orbits are determined by those that have apocenter radii greater than the bar radius.
%it is $ r_{\mathrm{apo}} > 4 \hspace{0.05cm} \mathrm{kpc} $ for model A and $ r_{\mathrm{apo}} > 13 \hspace{0.05cm} \mathrm{kpc} $ for model B.  
Then the orbits within the bar are classified into:   \textbf{\textit{boxo}} (ordinary box), \textbf{\textit{boxp}} (periodic or resonant box, i.e. boxlet), \textbf{\textit{ztub}} (z-axis tube), \textbf{\textit{ztup}} (periodic or resonant z-tube), \textbf{\textit{xtub}} (x-axis tube), \textbf{\textit{xtup}} (periodic or resonant x-tube), \textbf{\textit{x2++}} (similar to z-axis tube but elongated along the y-axis of the bar and prograde), \textbf{\textit{x4++}} (similar to z-axis tube but elongated along the y-axis and retrograde), \textbf{\textit{bo32}} (box with 3:2 fish resonance), \textbf{\textit{bobr}} (box brezel with 5:3 resonance), \textbf{\textit{x1++}} (resonant or near resonant $x_{1}$ boxlet), \textbf{\textit{x1cl}} ($x_{1}$ with 3:1 resonance), \textbf{\textit{x122}} ($x_{1}$ with 2:2 resonance), \textbf{\textit{x132}} ($x_{1}$ with 3:2 resonance), \textbf{\textit{x1bn}} ($x_{1}$ banana). We lumped all of the periodic $x_{1}$ resonant orbits together since there are very few.
\par
\cite{Valluri.2016} demonstrated that this automated orbit classification is generally consistent with visual classification by examining 20,000 individual orbits. The differences in classification accuracy range from approximately $ \sim 1\%$ to $ \sim 4 \% $ depending on the types of orbits being classified.
\par
NAFF employs the frequency drift parameter (diffusion rate) to measure the fraction of chaotic orbits. The frequency drift parameter $ \log_{10}(\Delta f) $ is computed by the change of the frequency in two equal orbital time segments \cite[]{Valluri2.2010}.
We use $\log_{10} (\Delta f) >-1.2$, which is a good empirical criterion for determining chaotic orbits as tested in \cite{Valluri2.2010} and \cite{Behzad.2021}. All orbits classified into the aforementioned families are additionally tagged with \textbf{\textit{regl}} (regular) or \textbf{\textit{chao}} (chaotic).
\par
We found that some orbits initially classified as periodic boxes exhibit chaotic behavior after a short time. Consequently, those orbits demonstrating chaotic characteristics were reclassified from the resonant orbit classes  \textbf{\textit{boxp}}, \textbf{\textit{bo32}}, and \textbf{\textit{bobr}} to non-periodic box class \textbf{\textit{boxo}}. Furthermore, we noted that some orbits initially classified as periodic z-axis tubes (\textbf{\textit{ztup}}), characterized by $\overline{L}_{zn} < 0.0$ and $-0.85 < \overline{L}_{xn} < 0.85$, should be reclassified into the \textbf{\textit{x4++}} group. Similarly, those with $\overline{L}_{zn} < 0.0$ and $\overline{L}_{xn} \sim \pm 1$ are more accurately categorized as \textbf{\textit{xtub}} orbits. Although these adjustments result in minor changes to orbital classes, they are crucial for our subsequent analysis of the morphology of different orbit classes.

\subsection{Orbital decomposition of CX and OX structure}
We further categorize the orbit types mentioned above by grouping together orbital families that exhibit similar morphologies.

\par
\textbf{(1) OX}: These are prograde short-axis tube orbits that are elongated along the bar, including 
 \textbf{\textit{x1++}}, \textbf{\textit{x1cl}}, \textbf{\textit{x122}}, \textbf{\textit{x132}}, \textbf{\textit{x1bn}}, \textbf{\textit{ztub}}, and \textbf{\textit{ztup}}. They make up an OX structure.
\par
\begin{figure*}
	\centering	%
	\includegraphics[width=0.9\linewidth]{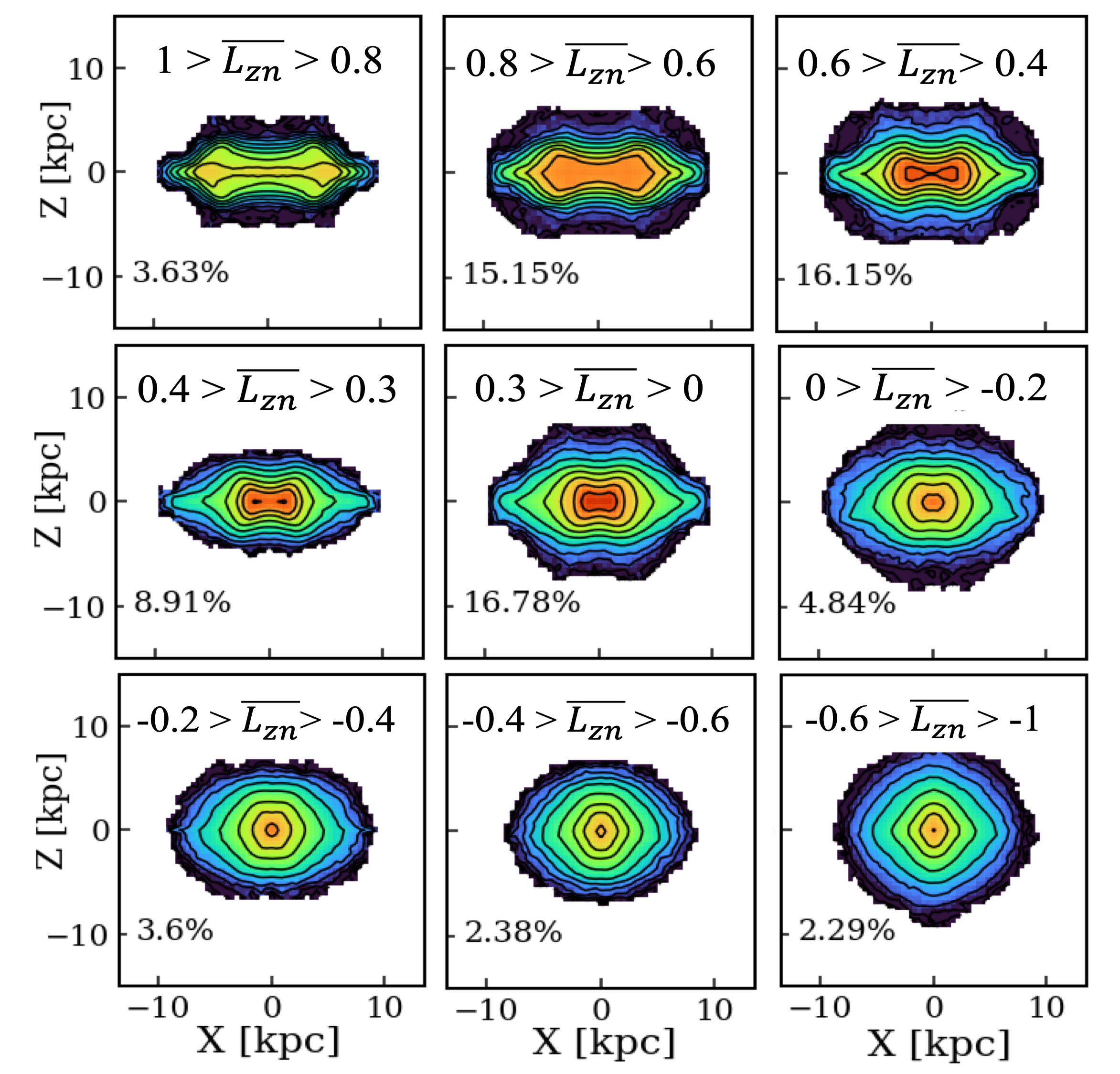}
	\caption{Projected surface density derived from box orbits in Model B at $t=4.5$ Gyr. Sequential panels display the dependence of box orbit morphologies across bins of the net rotation parameter $\overline{L}{zn}$, with the fractional contribution of orbits in each $\overline{L}{zn}$ bin to the total structure given as a percentage. Box orbits with higher normalized angular momentum   $1>\overline{L}{zn}>0.3$, predominantly facilitate the OX structure, whereas those within the range of $0.3>\overline{L}{zn}>-0.2$ are forming the CX structure.}%
	\label{fig:X_boxes}%
\end{figure*}

\textbf{(2) CX}: These are box orbits that are elongated along the bar, contributing to the formation of the CX structure. To select these groups of orbits, we begin by aggregating both periodic and non-periodic box orbits, including  \textbf{\textit{boxo}},  \textbf{\textit{boxp}}, \textbf{\textit{bo32}}, and \textbf{\textit{bobr}}. Box orbits exhibit a diverse range of morphologies. We have determined that the parameter $\overline{L}_{zn}$ serves as an effective criterion for distinguishing between box orbits that support a CX shape and those that exhibit a more rounded boxy morphology. Fig. \ref{fig:X_boxes} presents the surface densities of box orbits across various $\overline{L}_{zn}$ bins. The columns from left to right display the surface density in the $x-z$ planes for different $\overline{L}_{zn}$ bins, which is plotted for model B at $t=4.5$ Gyr.

Box orbits possessing higher values of $\overline{L}_{zn}$ prominently exhibit an OX structure, which transitions to a CX shape as the value decreases. With further reduction in $\overline{L}_{zn}$, the X wings gradually disappear. By examining the structures within various bins, we have identified a specific value of $\overline{L}_{zn}$ to serve as a threshold for visually distinguishing between OX and CX structures across all box orbits. For Model B, as depicted in Fig. \ref{fig:X_boxes}, box orbits with $\overline{L}_{zn} > 0.3$ are categorized as the OX group as they clearly exhibit a bridge connecting the two X wings. Those with $0.3 > \overline{L}_{zn} > -0.2$ are classified as the CX group, while the remaining orbits, which form a round shape, are included in the general box orbit group. These boundaries are determined visually and may vary from model to model. It should be noted that although altering the threshold value of $\overline{L}_{zn}$ can result in minor variations in the contribution of OX/CX orbits, these changes are typically limited to a few percent and are not significant.

\par
\textbf{(3) box}: These orbits are dominated by retrograde motion and are not included in CX or OX groups.
\par
\textbf{(4) x}$_{4}$: These orbits are retrograde short-axis tube orbits elongated perpendicular to the bar, referred to as \textbf{\textit{x4++}} in the NAFF classification. It is worth noting that \textbf{\textit{x2++}} orbits are their prograde counterparts, also short-axis tubes with a similar perpendicular orientation. However, our models do not contain any \textit{x2} orbits.
\par
\textbf{(5) LAT}:  These are long-axis tube orbits - orbits that have net angular momentum about the long $x$-axis of the bar model.
\par
\textbf{(6) disk}: Orbits with apocenter radii exceeding half the length of the bar are classified as disk orbits.

\section{Results}
\label{Xshape}
\begin{figure*}
	\includegraphics[trim=12 0 0 0, clip, width=\linewidth]{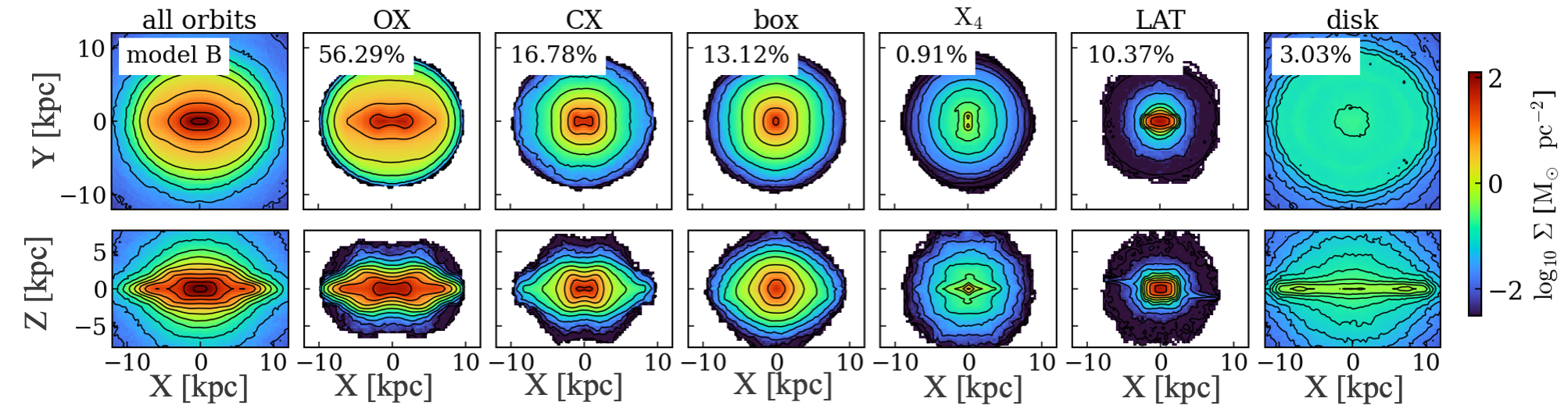}
	\caption{Projected surface density from 15,000 selected orbits in Model B at $t=4.5$ Gyr. The first row displays the $x-y$ plane, while the second row shows the $x-z$. The columns from left to right, represent the surface densities of all orbits, OX, CX, box, $x_{4}$, LAT, and disk orbits, respectively. Each panel is labeled with the fractional contribution of each orbital structure to the total, given as a percentage.} %
	\label{fig:all_den}%
\end{figure*}

Fig. \ref{fig:all_den} presents the extracted surface densities in the $ x - y $ and $ x - z $ planes for six orbit categories. The columns, from left to right, display surface densities constructed from orbits in the OX, CX, box, $x_{4}$, LAT, and disk categories, respectively. In all models, the OX and CX orbits contribute to structures elongated along the bar, forming OX- and CX-shaped structures respectively (shown in the 2nd and 3rd columns). The box, $x_{4}$, and LAT orbits do not contribute to the BP/X shape, instead supporting rounder or slightly boxy structures. In the following, we explore the orbital origins and evolution of the OX and CX structures in detail.

Furthermore, we demonstrate that orbits supporting the OX/CX vertical structures contribute to outer/central bar-elongated structures when the disk is viewed face-on. Our results confirm the lack of a relationship between X-shaped structures and the inner rounded parts of the face-on bar (e.g., barlenses). \cite{Beraldo.2023} illustrated the importance of orbits with loops at their ends (in the $x-y$ plane) to the bar shoulders in face-on density profiles. Additionally, they showed that bar thickening and vertical resonances can dilute the shoulders in the bar major-axis density profiles. We confirm that it is the OX rather than the CX orbits that contribute to the bar shoulders. This conclusion aligns with the findings from \cite{Beraldo.2023}, since the OX orbits have higher angular momentum and are therefore closer to their parent $x_{1}$ orbits with loops at their ends, while the CX orbits have lower angular momentum and are vertically thicker.

\subsection{Orbital origin of CX and OX structure}

\begin{figure*}% 
	\includegraphics[width=1\linewidth]{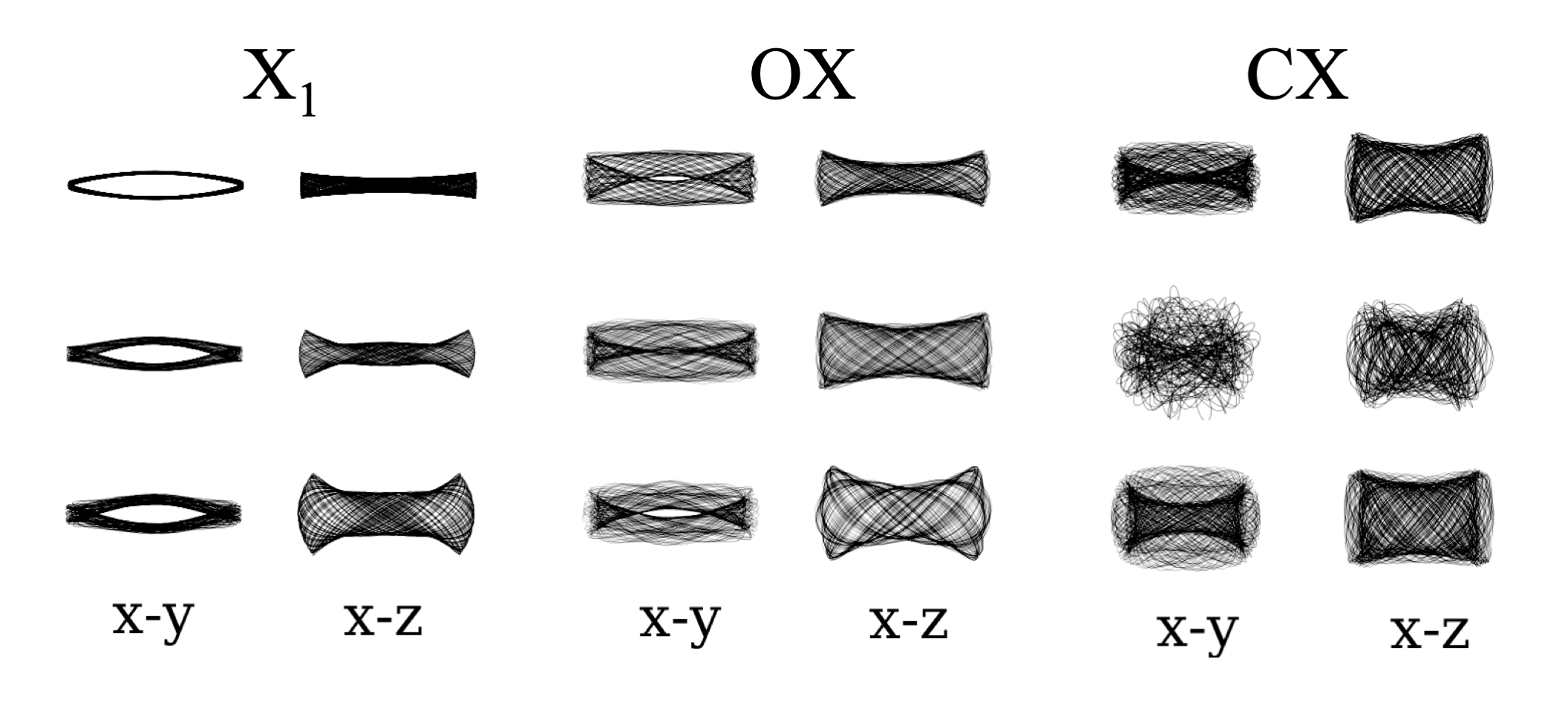}
	\caption{Illustration of typical $x_{1}$, OX, and CX orbits in the $x-y$ and $x-z$ planes.}%
 \label{fig:Xorb} 
\end{figure*}

The fraction of periodic orbits, including closed $x_{1}$ orbits, is less than $<3\%$ in all our models, and the majority of orbits that make up the BP/X-shaped structure are non-periodic. This is consistent with previous studies, which show that periodic orbits (such as banana, pretzel, etc.) constitute only a small fraction of orbits in N-body bars. Fig. \ref{fig:Xorb} depicts typical $x_{1}$ orbits (left panel), non-periodic orbits generating the OX (middle panel), and CX structures (right panel) in the $x-y$ and $x-z$ planes.

\cite{Parul.2020} argued that assembling non-periodic orbits into a pattern resembling a `bowtie' can create X-structures. These orbits occupy large areas on the $x-z$ plane and can be linked to quasi-periodic orbits surrounding the plane of the $x_{1}$ orbital family. For examples of such orbits, see Fig. 15 in \cite{Parul.2020}, Fig. 2 in \cite{Patsis.2014a}, and Fig. 9 in \cite{Patsis.2014b}. 

Based on the computed frequencies from NAFF, we confirm that the orbits contributing to the OX and CX structures can exhibit a wide range of $\Omega_{x}/\Omega_{z}$ values, extending beyond those typical of resonant orbits, such as banana or pretzel orbits. This finding is in agreement with previous studies.

As we demonstrated in Fig. \ref{fig:Xorb}, the typical non-periodic orbits within the OX and CX groups exhibit similar bowtie-like shapes. However, OX orbits tend to be vertically thinner bowties, whereas CX orbits are characterized by centrally thicker bowties. \cite{Parul.2020} suggested that further investigations using the Poincaré surfaces of section (SoS) are required to elucidate the origins of such orbits.%these orbits have a ratio of frequencies $ \Omega_{x}/\Omega_{z} $ in range of $0.55-0.62$. 
%that additional studies on the Poincare surface of sections are necessary to justify the origin of such orbits. 
%\par In the next section, we will explore the range of frequencies ratio $ f_{z}/ f_{x} $ of such orbits and investigate their origin using Poincare surface of sections plots.
\par
To elucidate the origin and characteristics of orbits within each OX and CX structure, in the following, we examine their phase-space and investigate their origin using Poincar\'e SoS plots.

\subsubsection{Poincar\'e surface of section}

A useful technique for exploring the distribution of orbits in phase space is to visualize their Poincar\'e SoS. Typically, the SoS is plotted for a set of orbits that share the same Jacobi integral value $ \mathrm{E_{J}} $. For orbits in two-dimensional bars, the SoS is plotted by mapping the velocity component $V_{y}$ against $y$ at each time step when an orbit intersects the $x$-axis with a negative $V_{x}$ value \cite[]{Sellwood.1993}.
 
\par

Plotting the SoS of orbits in an N-body simulation is complicated and presents two challenges: firstly, since all the orbits possess three-dimensional structures, they cannot be adequately represented on a two-dimensional SoS. Secondly, the distribution of $\mathrm{E_{J}}$ values is broad and continuous (rather than discrete), leading to fuzzy curves on the SoS. However, \cite{Shen.2004} demonstrated that a two-dimensional SoS can still offer valuable insights into the underlying three-dimensional orbital structures. 
%They achieved this by focusing on time steps when orbits are confined to the equatorial (i.e., $x-y$) plane, rather than the $x$-axis \vw{ I am confused by "confined to .... x axis"}.

\par
\begin{figure*}[t]
	\centering	%
	\includegraphics[width=\linewidth]{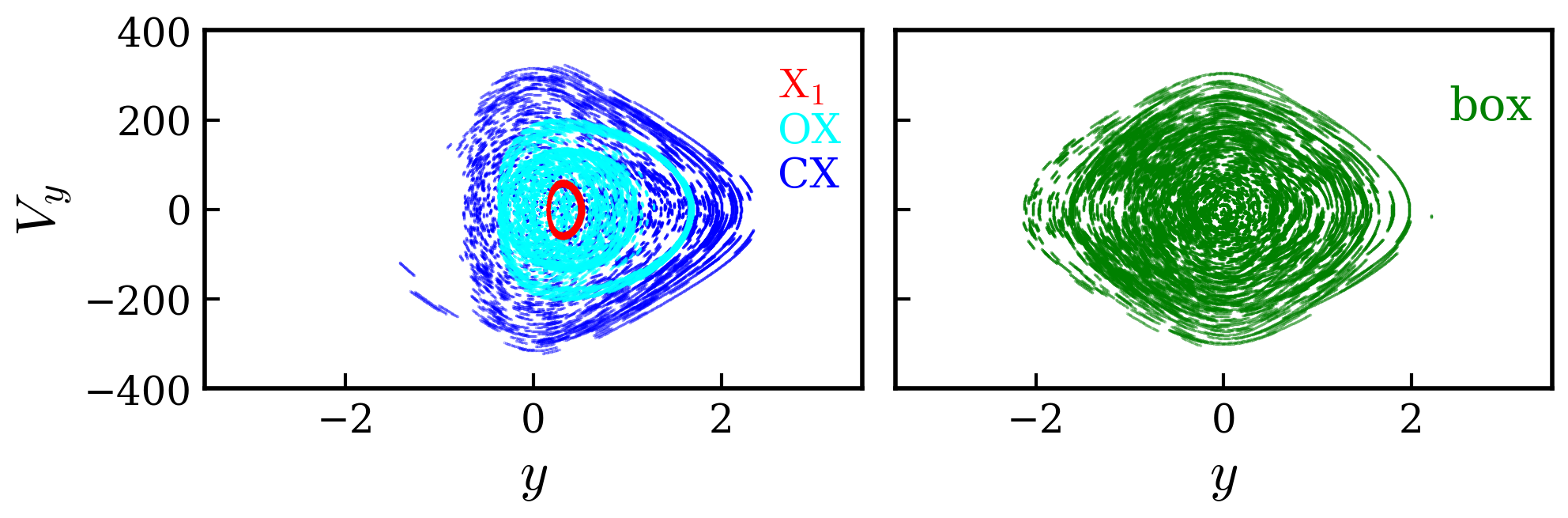}
	\caption{Surfaces of Section (SoS) for 50 orbits in model B at $t=4.5$ Gyr within the energy range $-1.0 < E_{J} < -0.98$. Each panel displays $V_{y}$ versus $y$ for orbits intersecting the $y-z$ plane with positive $V_{x}$. The colors denote various orbit families: $x_{1}$ orbits in red, OX orbits in cyan, CX orbits in blue, and box orbits in green. To alleviate overcrowding, the SoS for box orbits is plotted in the right panel. }%
	\label{fig:SoS_all}%
\end{figure*}

We explore the SoS of $100$ orbits in our models. Fig. \ref{fig:SoS_all} illustrates $V_{y}$ versus $y$ when orbits cross the $y-z$ plane with a positive $V_{x}$. This is plotted for orbits within the energy range of $-1.0 < E_{J} < -0.98$ in Model B at $t=4.5$ Gyr. The spread in $\mathrm{E_{J}}$ values results in fuzzy curves in the SoSs, particularly when plotted for a large number of orbits, as highlighted in Fig. 10 of \cite{Valluri.2016}. In our analysis of frozen potentials, most orbits traverse a broader area rather than being confined to narrow curves in the SoSs.

\par
The left panel of Fig. \ref{fig:SoS_all} displays the SoS for $x_{1}$ orbits (in red), OX orbits (in cyan), and CX orbits (in blue). To avoid overcrowding, the SoS for box orbits (in green) is presented separately in the right panel. The $ x_{1} $ orbits are the ovals at the center of the bull's eye of the SoS. It is similar to what is shown in  \cite{Shen.2004} (see Fig. 9) and \cite{Valluri.2016} (see Fig. 10). OX orbits are primarily clustered around $V_{y} = 0$ at positive $y$ values, confirming they are parented by $x_{1}$ orbits and exhibit prograde motion around the $z$ axis. Orbits in the red and cyan regions contribute to the formation of the OX structure. CX orbits, while also lying on oval curves surrounding the $x_{1}$ orbits, cover a wider region, indicating their inclusion in the same sequence. CX orbits are predominantly found at positive $y$ values. However, as CX orbits identified with $0.3>\overline{L}_{zn}>-0.2$, SoS of those orbits with negative $L_{z}$ encompass negative $y$ values. Box orbits lack a dominant direction of rotation, implying that as orbits evolve into a box-like shape, their angular momentum approaches zero on average.

\par
In summary, we confirm that orbits closer to the $x_{1}$ family exhibit an OX structure. As they move away from the $x_{1}$ orbits and become thicker, they transition to a CX shape. Eventually, they evolve into completely boxy or rounded shapes as they lose angular momentum.
%If the bar does not undergo a buckling process, as in model A, these transitions can happen hierarchically.
\par

%llustrates $V_{y}$ versus $y$ when orbits cross the $y-z$ plane with a positive $V_{x}$. This is plotted for obits within the energy range of $-1.0 < E_{j} < -0.98$ in Model B at $t=4.5$ Gyr.

\begin{figure*}
	\centering	%width=2\columnwidth
	\includegraphics[width=\linewidth]{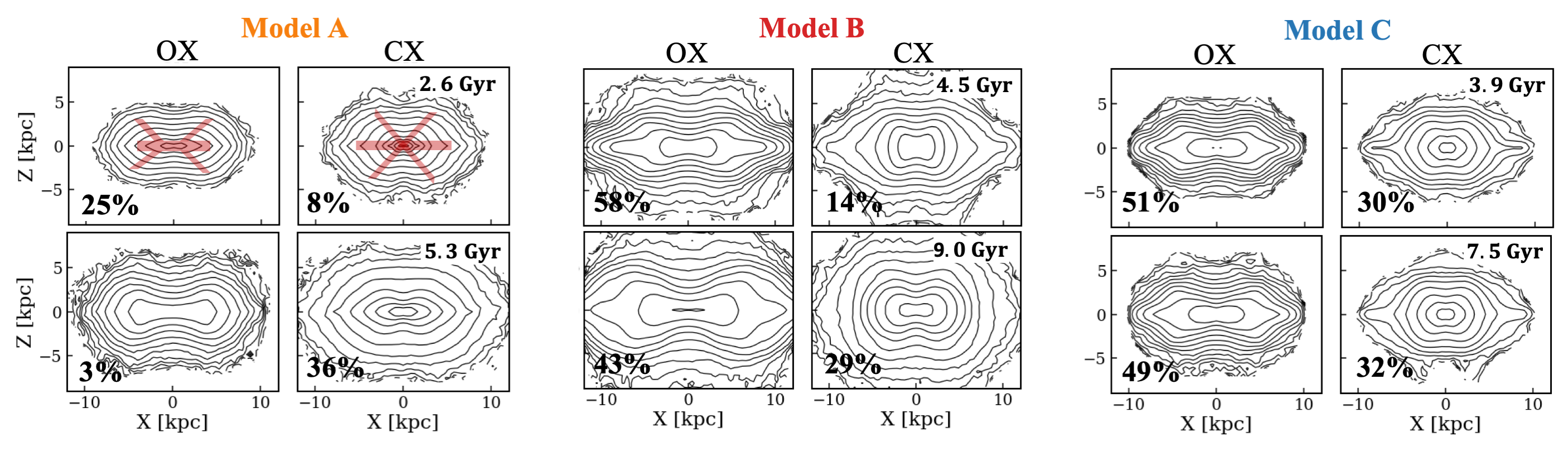}
	\hspace{8pt}%
	\caption{Contours of density extracted from OX and CX orbits and their contributions to models A, B, and C in edge-on views are presented. Red lines delineate various types of X-shaped structures. The first row represents the contributions of OX/CX structures after the formation of the BP/X bulge, while the second row displays these structures at the end of the simulation, with the contribution of each species to the total structure given as a percentage.}%
 %Left column for each model: the OX structures, in which the wings of X do not cross the bulge center, is supported by OX orbits in both models. Right column: the CX structure, in which the wings of X cross the center of the bulge and is supported by CX orbits in both models.  }%
	\label{fig:oxcx}%
\end{figure*}

\subsection{Evolution of CX and OX orbital structures}
Fig. \ref{fig:oxcx} presents contour densities of OX and CX orbital structures in model A (left), model B (middle), and model C (right). The first row displays these orbital structures shortly after the formation of the BP/X structure, while the second row illustrates these structures at the end of the simulations. 
\par
The proportions of OX and CX orbits in the models at various times are indicated in Fig. \ref{fig:oxcx}. In Model A, the BP/X bulge does not undergo buckling, and its pattern speed decreases significantly over time. Initially, the dominance of the OX orbits is evident at $25\%$ of all orbits, while the CX orbits represent only $8\%$. By the end of the simulation, the OX orbits have almost completely disappeared, dropping to $3\%$. In contrast, the BP/X bulge becomes predominantly characterized by the CX orbits, which increase to $36\%$.

Model B and Model C both undergo buckling, leading to the formation of the BP/X bulge where both OX and CX orbits coexist. In Model B, as the bar experiences a slowdown by the end of the simulation, the proportion of OX orbits decreases while that of CX orbits increases. The first row in the middle panel shows the OX/XC structure of model B after the first buckling event, and the bottom row displays the OX/XC structure after the second buckling. In contrast, in Model C, where the bar pattern speed remains constant, the proportions of CX and OX orbits stay the same by the end of the simulation. This suggests that as long as the bar's pattern speed does not decrease, the proportions of CX and OX orbits, and therefore the morphology of the BP/X structure remain unchanged. 

We defer to a future study the exploration of whether there is a relationship between the intensity of buckling and the proportions of CX and OX orbits after the formation of the BP/X structure.

\subsection{Photometric Parametrization of BP/X Bulges}

We use the method presented by \cite{Dattathri.2023} and IMFIT software \cite{Erwin.2015} to parameterize the shape of the BP/X bulges. In this parametrization, the bar is modeled using a $\rm sech^{2}$ profile similar to \citet{Picaud.2004, Robin.2012}, applied to a dimensionless, scaled radius given by
\begin{equation}
\rho  = \rho_0 \operatorname{sech}^2\left(-R_s\right),
\end{equation}
where
\begin{equation}
R_s=\left(\left[\left(\frac{x}{X_{\mathrm{bar}}}\right)^{c_{\perp}}+\left(\frac{y}{Y_{\mathrm{bar}}}\right)^{c_{\perp}}\right]^{c_{\|} / c_{\perp}}+\left(\frac{z}{Z_{\mathrm{bar}}}\right)^{c_{\|}}\right)^{1 / c_{\|}} .
\end{equation}
The coordinate system is centered at the galaxy's nucleus, with $ \rm X_{bar}$,
$\rm  Z_{bar}$ and $\rm  Y_{bar}$ denoting the semi-major, intermediate, and semi-minor axis lengths of the bar, respectively. $c_{\|}$ and $c_{\perp}$ control the diskiness/boxiness of the bar introduced by \cite{Athanassoula.1990}. Here, we adopt 
$c_{\|},c_{\perp} \in [1.5,5]$, based on the results of \cite{robin2012}. The BP/X feature's morphology is defined by a scale height perpendicular to the disk plane, which varies across the $x - y$  plane and is modeled by a double Gaussian distribution centered on the galactic center:
\begin{equation}
\begin{aligned}
Z_{\text {bar }}(x, y)= & A_{\text {pea }} \exp \left(-\frac{\left(x-R_{\text {pea }}\right)^2}{2 \sigma_{\text {pea }}^2}-\frac{y^2}{2 \sigma_{\text {pea }}^2}\right) + \\
& A_{\text {pea }} \exp \left(-\frac{\left(x+R_{\text {pea }}\right)^2}{2 \sigma_{\text {pea }}^2}-\frac{y^2}{2 \sigma_{\text {pea }}^2}\right)+z_0,
\end{aligned}
\end{equation}
where $\rm R_{\mathrm{pea}}$ and $\sigma_{\mathrm{pea}}$ denote the distance of the peanut's center from the galactic center and the width of each peanut, respectively. $\rm A_{\mathrm{pea}}$ quantifies the vertical extent of the peanut feature above the ellipsoidal bar's scale height $z_{0}$. 
This expression closely resembles the "peanut height function" described by \cite{Fragkoudi.2015}, with the distinction that it imposes symmetry between the two halves of the peanut relative to the galactic center and enforces alignment along the bar major axis.

As demonstrated by \cite{Dattathri.2023}, these three parameters provide substantial flexibility in capturing the diverse morphologies exhibited by the BP/X feature. Although these values do not directly quantify the OX/CX structures, they are useful for demonstrating the relationship between the evolution of BP/X morphology and the bar pattern speed independent of orbital analysis. The quantification of OX/CX structures through photometric parameterization will be addressed in future studies. 

Figure \ref{fig:Xparam} displays the variation of BP/X shape parameters $R_{\mathrm{pea}}$, $\sigma_{\mathrm{pea}}$, and $R_{\mathrm{bar}}$ versus the bar pattern speed (left column) and versus the dimensionless bar rotation parameter $ \mathcal{R} \equiv R_{\rm cor}/R_{\rm bar}$ (right column) for model A (orange), model B (red), and model C (blue). The $x$-axis of the left column is the bar pattern speed in the reverse direction. Models A and B exhibit significant changes in BP/X morphology over time as the bar pattern speed decreases. In contrast, the BP/X parameters in model C are relatively constant or change very little, which is attributed to its nearly constant bar pattern speed.
In models A and B, the relation between the BP/X parameters and the bar pattern speed is monotonic (although not linear). This holds true even after the second buckling in model B (represented by the red point number 7).
As the bar pattern speed decreases, all three parameters $R_{\mathrm{pea}}$, $\sigma_{\mathrm{pea}}$, and $R_{\mathrm{bar}}$  increase. However, we do not find a monotonic relationship between the BP/X parameters and  $\mathcal{R}$. %This could be because the bar parameters are sometimes subject to short-term oscillations, e.g., by bar-spiral interactions \cite{Hilmi.2020}. 
This indicates that the evolution of BP/X morphology is correlated with the evolution of the bar pattern speed rather than $\mathcal{R}$.

%\vw{It might be illuminating for the reader to relate the evolution of morphology quantified by these values to what morphology we see from the OX and CX orbits.}

\begin{figure*}%
	\centering
	\subfigure{ 
		\label{fig:Xparam} 
		\includegraphics[width=.5\linewidth]{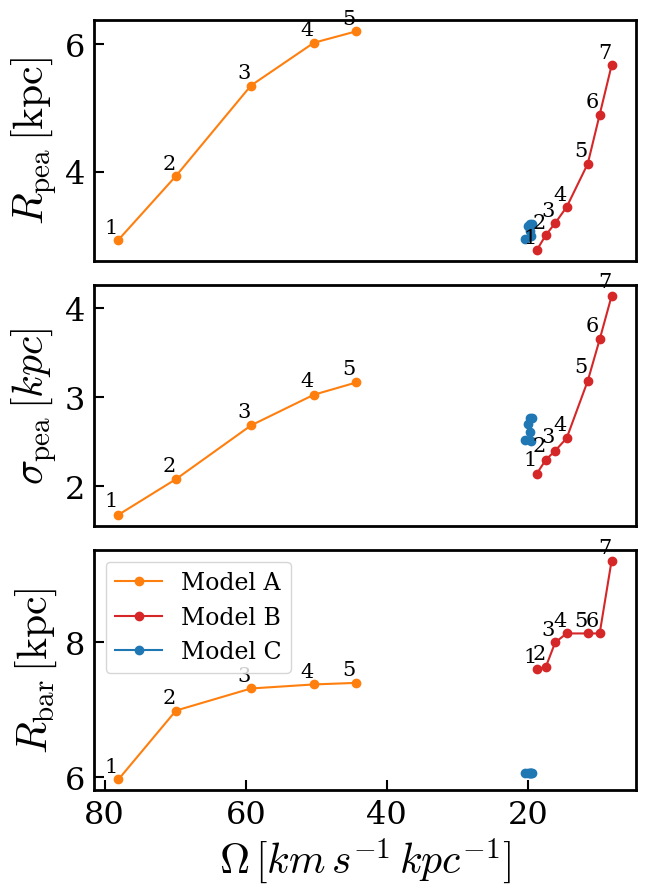}
  \includegraphics[width=.505\linewidth]{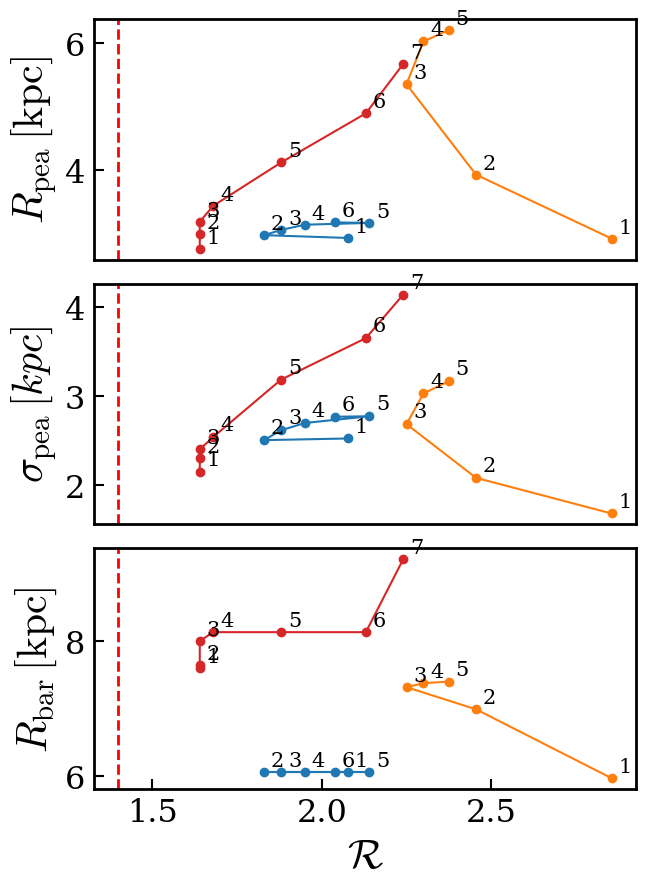}}
	\caption{Left column: The variation of BP/X shape parameters $R_{\mathrm{pea}}$ (top), $\sigma_{\mathrm{pea}}$ (middle), and  $R_{\mathrm{bar}}$ (bottom) with the bar pattern speed for model A (orange), model B (red), and model C (blue) over time. Note that the bar pattern speed for model A here is not scaled as it is in Figure \ref{fig:model_param}. The x-axis shows the pattern speed in the inverse direction. We use only snapshots where the bar is in a relatively steady state. Right column: The variation of BP/X shape parameters with the dimensionless bar rotation parameter $ \mathcal{R} \equiv R_{\rm cor}/R_{\rm bar}$. The vertical red dashed line indicates $\mathcal{R}=1.4$, the criterion below which the bar is classified as a fast bar. For model C, the  $\mathcal{R}$ values are multiplied by a factor of 0.5 for better visualization. The numbers tagged to each point denote the sequence of time evolution, ranging from 1 (earliest snapshot) to the latest snapshot.}
 %\vw{Make sure that the color labeling in this figure matches that of the other figures! The time units as well. I do not know what the simulation time units are for model A, but I assume these measurements are taken after the bulge has become prominent for each model since it does not start at 0 time. If so, mention it.} \vw{This is also a bit small, but since it is mainly qualitative, I might keep it basically as is, but reduce the number of ticks so the axis label font can be larger.} \mvco{fix the units of $R_{pea}, \sigma_{pea}$ and also the units of the x-axis should be time. May be worth plotting Rpea, sigpea against bar pattern speed (just a scatter plot for all values since for model C they will all overlap.}}%		
	\label{fig:Xparam}%
\end{figure*}

\subsection{OX/CX bulges and the bar pattern speed in Auriga simulations}
The Auriga simulations \cite[]{Grand.2017} are a suite of 30 magneto-hydrodynamical cosmological zoom-in simulations. The Auriga barred galaxies exhibit fast bars primarily because they are baryon-dominated disks, experiencing less dynamical friction, which prevents the bars from slowing down \cite[]{Fragkoudi.2021}.

\cite{Guillermo.2020} analyzed the structural and photometric properties of 21 barred galaxies from the Auriga simulations. Using edge-on unsharp-masked images at $z=0$, they identified BP/X structures in the inner parts of 6 out of 21 galaxies. They visually determined whether these BP/X structures are OX or CX, and found that only one of these bulges has CX morphology, while the other five have OX structures. See Table 3 in \cite{Guillermo.2020}.

\cite{Fragkoudi.2021} presented the evolution of the pattern speed for five Auriga galaxies with high cadence outputs as a function of lookback time. See Fig B.1 in \cite{Fragkoudi.2021}. Three of these galaxies, A17, A18, and A26, are identified as having BP/X bulges by \cite{Guillermo.2020}. A18 galaxy, the only one with a CX structure, has a relatively lower pattern speed of around  $27$ $\rm km s^{-1} kpc^{-1}$. A17 and A26, which have OX structures, exhibit significantly higher pattern speeds around $45$ and $37$ $\rm km s^{-1} kpc^{-1}$, respectively. Our results explain the findings in the Auriga simulations: bars with higher pattern speeds tend to exhibit BP/X bulges with more dominant OX structures. %evidencing a strong correlation between BP/X morphology and bar pattern speed.

\section{SUMMARY}
\label{conl}
%We use the auto-classification method in NAFF software based on the orbit frequency analysis. We analyzed the orbits in two N-body bar models with BP/X structure, and the main results are as following: %uses frequency ratios, time-averaged normalized angular momentum, apocenter radius, $y_{max}$, and $x_{max}$. 
We present a methodology that enables us to decompose the orbital structures supporting OX and CX shapes in BP/X bulges. Our method relies on auto-classification with NAFF software, focusing on the morphological and kinematic characteristics of orbits.

To understand the origin and evolution of OX and CX structures in BP/X bulges, we applied our method to classify orbits using a frozen potential from three N-body bar models, each with distinct features, at two different times, beginning shortly after the formation of the BP/X bulge. The main results are as follows:
\par
(1) We demonstrate that both OX and CX structures are composed of non-periodic orbits associated with the $x_{1}$ orbit family. OX orbits, being closer to $x_{1}$ orbits, possess higher angular momentum compared to those in CX structures. 
\par
(2) We found that orbits supporting OX/CX structures in the edge-on view also form a similar OX/CX-like shape in the face-on view. 
\par
(3) We showed that as the bar pattern speed decreases, the contribution from OX orbits decreases while that from CX orbits increases. If the bar remains at a constant speed, the shape of the BP/X structure stays unchanged until the end of the simulation. This indicates a strong dependence of the BP/X bulge shape on the bar pattern speed.
\par
(4) We show that Model A (the bar that does not buckle) is initially dominated by OX orbits. These orbits surrounding the $x_{1}$ orbits begin to thicken. As the bar loses angular momentum, the OX structure transitions to a CX structure by the end of the simulation. In contrast, our bar models that experience buckling events exhibit a combination of both OX and CX structures immediately after the buckling and evolve toward CX dominance if the pattern speed decreases.
\par
(5) Using photometric parameterization of the BP/X structure, we find that the evolution of BP/X morphology is correlated with the bar pattern speed rather than the bar rotation parameter.

%(5) \mvco{Is it worth adding the point that the fact that OX shapes are associated with bars with high angular momentum and CX with bars with low angular momentum means that images of edge-on bars alone could provide a way to infer bar pattern speeds and angular momenta? Maybe too speculative but I think it is potentially interesting.}

%(5) This approach is applicable to decompose orbital structures of BP/X bulges in N-body bars, orbit-based Galactic Bulge studies using surveys, Made-to-Measure, and Schwarzschild dynamical models.

\section*{Acknowledgements}
We thank J. A. Sellwood for providing the data for Model A. BT, SD, MV, VW, LBeS gratefully acknowledge funding from the National Science Foundation (grant NSF-AST-2009122 to MV). LBeS is supported by the Heising Simons Foundation through the Barbara Pichardo Future Faculty Fellowship from grant \# 2022-3927.

%The research presented here is partially supported by the National Key R\&D Program of China under grant No. 2018YFA0404501; by the National Natural Science Foundation of China under grant Nos. 12025302, 11773052, 11761131016; by the ``111'' Project of the Ministry of Education of China under grant No. B20019; and by the Chinese Space Station Telescope project. This work made use of the facilities of the Center for High-Performance Computing at Shanghai Astronomical Observatory. LZ acknowledges the support from National Natural Science Foundation of China under grant No. Y945271001.

\textit{Software}: \textit{AGAMA} \cite[]{Vasiliev.2019}, \textit{NAFF} \cite[]{Valluri.1998, Valluri.2016}, \textit{Jupyter Notebook} \cite[]{Kluyver.2016},  \textit{matplotlib} \cite[]{Hunter.2007}, \textit{numpy} \cite[]{Harris.2020},  \textit{scipy} \cite[]{Virtanen.2020}.  

%%%%%%%%%%%%%%%%%%%%%%%%%%%%%%%%%%%%%%%%%%%%%%%%%%
\section*{Data Availability}
The simulation data for model B and model C are available at doi:\dataset[10.5281/zenodo.11237062]{https://doi.org/10.5281/zenodo.11237062} Readers wishing to obtain the simulation data for Model A should contact J. A. Sellwood. Any additional data will be shared on reasonable request to the corresponding author.
%\mvch{give Zenodo link of Vance's paper - Vance please add two snapshots of Run 6000 to that archive so that it is available.} \vw{I have added the doi link, which will point to the newest version. Behzad, please share with me the specific snapshot numbers for each model you used. Currently no data for run 6000 is in the repository, and only the initial and final snapshots for run 2000 are present, so I will need snap numbers from both.} 

\bibliographystyle{aasjournal}
\bibliography{Bibtex}

\begin{thebibliography}{}
\expandafter\ifx\csname natexlab\endcsname\relax\def\natexlab#1{#1}\fi
\providecommand{\url}[1]{\href{#1}{#1}}
\providecommand{\dodoi}[1]{doi:~\href{http://doi.org/#1}{\nolinkurl{#1}}}
\providecommand{\doeprint}[1]{\href{http://ascl.net/#1}{\nolinkurl{http://ascl.net/#1}}}
\providecommand{\doarXiv}[1]{\href{https://arxiv.org/abs/#1}{\nolinkurl{https://arxiv.org/abs/#1}}}

\bibitem[{{Abbott} {et~al.}(2017){Abbott}, {Valluri}, {Shen}, \&
  {Debattista}}]{Abbott.2017}
{Abbott}, C.~G., {Valluri}, M., {Shen}, J., \& {Debattista}, V.~P. 2017,
  \mnras, 470, 1526, \dodoi{10.1093/mnras/stx1262}

\bibitem[{{Athanassoula} {et~al.}(1990){Athanassoula}, {Morin}, {Wozniak},
  {Puy}, {Pierce}, {Lombard}, \& {Bosma}}]{Athanassoula.1990}
{Athanassoula}, E., {Morin}, S., {Wozniak}, H., {et~al.} 1990, \mnras, 245, 130

\bibitem[{{Beraldo e Silva} {et~al.}(2023){Beraldo e Silva}, {Debattista},
  {Anderson}, {Valluri}, {Erwin}, {Daniel}, \& {Deg}}]{Beraldo.2023}
{Beraldo e Silva}, L., {Debattista}, V.~P., {Anderson}, S.~R., {et~al.} 2023,
  \apj, 955, 38, \dodoi{10.3847/1538-4357/ace976}

\bibitem[{{Binney} \& {Tremaine}(2008)}]{Binny.2008}
{Binney}, J., \& {Tremaine}, S. 2008, {Galactic Dynamics: Second Edition}
  (Princeton University Press)

\bibitem[{{Bl{\'a}zquez-Calero} {et~al.}(2020){Bl{\'a}zquez-Calero}, {Florido},
  {P{\'e}rez}, {Zurita}, {Grand}, {Fragkoudi}, {G{\'o}mez}, {Marinacci}, \&
  {Pakmor}}]{Guillermo.2020}
{Bl{\'a}zquez-Calero}, G., {Florido}, E., {P{\'e}rez}, I., {et~al.} 2020,
  \mnras, 491, 1800, \dodoi{10.1093/mnras/stz3125}

\bibitem[{{Bureau} {et~al.}(2006){Bureau}, {Aronica}, {Athanassoula},
  {Dettmar}, {Bosma}, \& {Freeman}}]{Bureau.2006}
{Bureau}, M., {Aronica}, G., {Athanassoula}, E., {et~al.} 2006, \mnras, 370,
  753, \dodoi{10.1111/j.1365-2966.2006.10471.x}

\bibitem[{{Carpintero} \& {Aguilar}(1998)}]{Carpintero.1998}
{Carpintero}, D.~D., \& {Aguilar}, L.~A. 1998, \mnras, 298, 1,
  \dodoi{10.1046/j.1365-8711.1998.01320.x}

\bibitem[{{Combes} {et~al.}(1990){Combes}, {Debbasch}, {Friedli}, \&
  {Pfenniger}}]{Combes.1990}
{Combes}, F., {Debbasch}, F., {Friedli}, D., \& {Pfenniger}, D. 1990, \aap,
  233, 82

\bibitem[{{Contopoulos} \& {Harsoula}(2013)}]{Contopoulos.2013}
{Contopoulos}, G., \& {Harsoula}, M. 2013, \mnras, 436, 1201,
  \dodoi{10.1093/mnras/stt1640}

\bibitem[{{Dattathri} {et~al.}(2024){Dattathri}, {Valluri}, {Vasiliev},
  {Wheeler}, \& {Erwin}}]{Dattathri.2023}
{Dattathri}, S., {Valluri}, M., {Vasiliev}, E., {Wheeler}, V., \& {Erwin}, P.
  2024, \mnras, 530, 1195, \dodoi{10.1093/mnras/stae802}

\bibitem[{{Debattista} {et~al.}(2020){Debattista}, {Liddicott},
  {Khachaturyants}, \& {Beraldo e Silva}}]{Debattista.2020}
{Debattista}, V.~P., {Liddicott}, D.~J., {Khachaturyants}, T., \& {Beraldo e
  Silva}, L. 2020, \mnras, 498, 3334, \dodoi{10.1093/mnras/staa2568}

\bibitem[{{Dehnen} {et~al.}(2023){Dehnen}, {Semczuk}, \&
  {Sch{\"o}nrich}}]{Dehnen.2023}
{Dehnen}, W., {Semczuk}, M., \& {Sch{\"o}nrich}, R. 2023, \mnras, 518, 2712,
  \dodoi{10.1093/mnras/stac3184}

\bibitem[{{Erwin}(2015)}]{Erwin.2015}
{Erwin}, P. 2015, \apj, 799, 226, \dodoi{10.1088/0004-637X/799/2/226}

\bibitem[{{Erwin} \& {Debattista}(2017)}]{Erwin.2017}
{Erwin}, P., \& {Debattista}, V.~P. 2017, \mnras, 468, 2058,
  \dodoi{10.1093/mnras/stx620}

\bibitem[{{Fragkoudi} {et~al.}(2015){Fragkoudi}, {Athanassoula}, {Bosma}, \&
  {Iannuzzi}}]{Fragkoudi.2015}
{Fragkoudi}, F., {Athanassoula}, E., {Bosma}, A., \& {Iannuzzi}, F. 2015,
  \mnras, 450, 229, \dodoi{10.1093/mnras/stv537}

\bibitem[{{Fragkoudi} {et~al.}(2021){Fragkoudi}, {Grand}, {Pakmor}, {Springel},
  {White}, {Marinacci}, {Gomez}, \& {Navarro}}]{Fragkoudi.2021}
{Fragkoudi}, F., {Grand}, R.~J.~J., {Pakmor}, R., {et~al.} 2021, \aap, 650,
  L16, \dodoi{10.1051/0004-6361/202140320}

\bibitem[{{Freudenreich}(1998)}]{Freudenreich.1998}
{Freudenreich}, H.~T. 1998, \apj, 492, 495, \dodoi{10.1086/305065}

\bibitem[{{Grand} {et~al.}(2017){Grand}, {G{\'o}mez}, {Marinacci}, {Pakmor},
  {Springel}, {Campbell}, {Frenk}, {Jenkins}, \& {White}}]{Grand.2017}
{Grand}, R. J.~J., {G{\'o}mez}, F.~A., {Marinacci}, F., {et~al.} 2017, \mnras,
  467, 179, \dodoi{10.1093/mnras/stx071}

\bibitem[{{Harris} {et~al.}(2020){Harris}, {Millman}, {van der Walt},
  {Gommers}, {Virtanen}, {Cournapeau}, {Wieser}, {Taylor}, {Berg}, {Smith},
  {Kern}, {Picus}, {Hoyer}, {van Kerkwijk}, {Brett}, {Haldane}, {del R{\'\i}o},
  {Wiebe}, {Peterson}, {G{\'e}rard-Marchant}, {Sheppard}, {Reddy}, {Weckesser},
  {Abbasi}, {Gohlke}, \& {Oliphant}}]{Harris.2020}
{Harris}, C.~R., {Millman}, K.~J., {van der Walt}, S.~J., {et~al.} 2020, \nat,
  585, 357, \dodoi{10.1038/s41586-020-2649-2}

\bibitem[{{Hunter}(2007)}]{Hunter.2007}
{Hunter}, J.~D. 2007, Computing in Science and Engineering, 9, 90,
  \dodoi{10.1109/MCSE.2007.55}

\bibitem[{Kluyver {et~al.}(2016)Kluyver, Ragan-Kelley, P{\'e}rez, Granger,
  Bussonnier, Frederic, Kelley, Hamrick, Grout, Corlay, Ivanov, Avila, Abdalla,
  \& Willing}]{Kluyver.2016}
Kluyver, T., Ragan-Kelley, B., P{\'e}rez, F., {et~al.} 2016, in Positioning and
  Power in Academic Publishing: Players, Agents and Agendas, ed. F.~Loizides \&
  B.~Schmidt, IOS Press, 87 -- 90

\bibitem[{{Li} {et~al.}(2017){Li}, {Ho}, \& {Barth}}]{Li.2017}
{Li}, Z.-Y., {Ho}, L.~C., \& {Barth}, A.~J. 2017, \apj, 845, 87,
  \dodoi{10.3847/1538-4357/aa7fba}

\bibitem[{{L{\"u}tticke} {et~al.}(2000){L{\"u}tticke}, {Dettmar}, \&
  {Pohlen}}]{Ltticke.2000}
{L{\"u}tticke}, R., {Dettmar}, R.~J., \& {Pohlen}, M. 2000, \aaps, 145, 405,
  \dodoi{10.1051/aas:2000354}

\bibitem[{{Merritt} \& {Sellwood}(1994)}]{Merritt.1994}
{Merritt}, D., \& {Sellwood}, J.~A. 1994, \apj, 425, 551,
  \dodoi{10.1086/174005}

\bibitem[{{Navarro} {et~al.}(1996){Navarro}, {Frenk}, \& {White}}]{NFW.1996}
{Navarro}, J.~F., {Frenk}, C.~S., \& {White}, S. D.~M. 1996, \apj, 462, 563,
  \dodoi{10.1086/177173}

\bibitem[{{Parul} {et~al.}(2020){Parul}, {Smirnov}, \&
  {Sotnikova}}]{Parul.2020}
{Parul}, H.~D., {Smirnov}, A.~A., \& {Sotnikova}, N.~Y. 2020, \apj, 895, 12,
  \dodoi{10.3847/1538-4357/ab76ce}

\bibitem[{{Patsis} \& {Athanassoula}(2019)}]{PatsisAna.2019}
{Patsis}, P.~A., \& {Athanassoula}, E. 2019, \mnras, 490, 2740,
  \dodoi{10.1093/mnras/stz2588}

\bibitem[{{Patsis} \& {Harsoula}(2018)}]{Patsis.2018}
{Patsis}, P.~A., \& {Harsoula}, M. 2018, \aap, 612, A114,
  \dodoi{10.1051/0004-6361/201731114}

\bibitem[{{Patsis} \& {Katsanikas}(2014{\natexlab{a}})}]{Patsis.2014a}
{Patsis}, P.~A., \& {Katsanikas}, M. 2014{\natexlab{a}}, \mnras, 445, 3525,
  \dodoi{10.1093/mnras/stu1988}

\bibitem[{{Patsis} \& {Katsanikas}(2014{\natexlab{b}})}]{Patsis.2014b}
---. 2014{\natexlab{b}}, \mnras, 445, 3546, \dodoi{10.1093/mnras/stu1970}

\bibitem[{{Patsis} {et~al.}(2002){Patsis}, {Skokos}, \&
  {Athanassoula}}]{Patsis.2002}
{Patsis}, P.~A., {Skokos}, C., \& {Athanassoula}, E. 2002, \mnras, 337, 578,
  \dodoi{10.1046/j.1365-8711.2002.05943.x}

\bibitem[{{Pfenniger} \& {Friedli}(1991)}]{Pfenniger.1991}
{Pfenniger}, D., \& {Friedli}, D. 1991, \aap, 252, 75

\bibitem[{{Picaud} \& {Robin}(2004)}]{Picaud.2004}
{Picaud}, S., \& {Robin}, A.~C. 2004, \aap, 428, 891,
  \dodoi{10.1051/0004-6361:20041218}

\bibitem[{{Portail} {et~al.}(2015){Portail}, {Wegg}, \&
  {Gerhard}}]{Portail.2015}
{Portail}, M., {Wegg}, C., \& {Gerhard}, O. 2015, \mnras, 450, L66,
  \dodoi{10.1093/mnrasl/slv048}

\bibitem[{{Quillen}(2002)}]{Quillen.2002}
{Quillen}, A.~C. 2002, \aj, 124, 722, \dodoi{10.1086/341753}

\bibitem[{{Quillen} {et~al.}(2014){Quillen}, {Minchev}, {Sharma}, {Qin}, \& {Di
  Matteo}}]{Quillen.2014}
{Quillen}, A.~C., {Minchev}, I., {Sharma}, S., {Qin}, Y.-J., \& {Di Matteo}, P.
  2014, \mnras, 437, 1284, \dodoi{10.1093/mnras/stt1972}

\bibitem[{{Raha} {et~al.}(1991){Raha}, {Sellwood}, {James}, \&
  {Kahn}}]{Raha.1991}
{Raha}, N., {Sellwood}, J.~A., {James}, R.~A., \& {Kahn}, F.~D. 1991, \nat,
  352, 411, \dodoi{10.1038/352411a0}

\bibitem[{{Robin} {et~al.}(2012{\natexlab{a}}){Robin}, {Marshall},
  {Schultheis}, \& {Reyl{\'e}}}]{Robin.2012}
{Robin}, A.~C., {Marshall}, D.~J., {Schultheis}, M., \& {Reyl{\'e}}, C.
  2012{\natexlab{a}}, \aap, 538, A106, \dodoi{10.1051/0004-6361/201116512}

\bibitem[{{Robin} {et~al.}(2012{\natexlab{b}}){Robin}, {Marshall},
  {Schultheis}, \& {Reyl{\'e}}}]{robin2012}
---. 2012{\natexlab{b}}, \aap, 538, A106, \dodoi{10.1051/0004-6361/201116512}

\bibitem[{{Sellwood}(2014)}]{Sellwood.2014}
{Sellwood}, J.~A. 2014, arXiv e-prints, arXiv:1406.6606,
  \dodoi{10.48550/arXiv.1406.6606}

\bibitem[{{Sellwood} \& {Gerhard}(2020)}]{Sellwood.2020}
{Sellwood}, J.~A., \& {Gerhard}, O. 2020, \mnras, 495, 3175,
  \dodoi{10.1093/mnras/staa1336}

\bibitem[{{Sellwood} \& {Wilkinson}(1993)}]{Sellwood.1993}
{Sellwood}, J.~A., \& {Wilkinson}, A. 1993, Reports on Progress in Physics, 56,
  173, \dodoi{10.1088/0034-4885/56/2/001}

\bibitem[{{Shaw}(1987)}]{Shaw.1987}
{Shaw}, M.~A. 1987, \mnras, 229, 691, \dodoi{10.1093/mnras/229.4.691}

\bibitem[{{Shen} \& {Sellwood}(2004)}]{Shen.2004}
{Shen}, J., \& {Sellwood}, J.~A. 2004, \apj, 604, 614, \dodoi{10.1086/382124}

\bibitem[{{Skokos} {et~al.}(2002){Skokos}, {Patsis}, \&
  {Athanassoula}}]{Skokos.2002}
{Skokos}, C., {Patsis}, P.~A., \& {Athanassoula}, E. 2002, \mnras, 333, 847,
  \dodoi{10.1046/j.1365-8711.2002.05468.x}

\bibitem[{{Tahmasebzadeh} {et~al.}(2021){Tahmasebzadeh}, {Zhu}, {Shen},
  {Gerhard}, \& {Qin}}]{Behzad.2021}
{Tahmasebzadeh}, B., {Zhu}, L., {Shen}, J., {Gerhard}, O., \& {Qin}, Y. 2021,
  \mnras, 508, 6209, \dodoi{10.1093/mnras/stab3002}

\bibitem[{{Valencia-Enr{\'\i}quez} {et~al.}(2023){Valencia-Enr{\'\i}quez},
  {Puerari}, \& {Chaves-Velasquez}}]{Valencia.2023}
{Valencia-Enr{\'\i}quez}, D., {Puerari}, I., \& {Chaves-Velasquez}, L. 2023,
  \mnras, 525, 3162, \dodoi{10.1093/mnras/stad2437}

\bibitem[{{Valluri} {et~al.}(2010){Valluri}, {Debattista}, {Quinn}, \&
  {Moore}}]{Valluri2.2010}
{Valluri}, M., {Debattista}, V.~P., {Quinn}, T., \& {Moore}, B. 2010, \mnras,
  403, 525, \dodoi{10.1111/j.1365-2966.2009.16192.x}

\bibitem[{{Valluri} \& {Merritt}(1998)}]{Valluri.1998}
{Valluri}, M., \& {Merritt}, D. 1998, \apj, 506, 686, \dodoi{10.1086/306269}

\bibitem[{{Valluri} {et~al.}(2016){Valluri}, {Shen}, {Abbott}, \&
  {Debattista}}]{Valluri.2016}
{Valluri}, M., {Shen}, J., {Abbott}, C., \& {Debattista}, V.~P. 2016, \apj,
  818, 141, \dodoi{10.3847/0004-637X/818/2/141}

\bibitem[{{Vasiliev}(2013)}]{Vasiliev.2013}
{Vasiliev}, E. 2013, \mnras, 434, 3174, \dodoi{10.1093/mnras/stt1235}

\bibitem[{{Vasiliev}(2019)}]{Vasiliev.2019}
---. 2019, \mnras, 482, 1525, \dodoi{10.1093/mnras/sty2672}

\bibitem[{{Virtanen} {et~al.}(2020){Virtanen}, {Gommers}, {Oliphant},
  {Haberland}, {Reddy}, {Cournapeau}, {Burovski}, {Peterson}, {Weckesser},
  {Bright}, {van der Walt}, {Brett}, {Wilson}, {Millman}, {Mayorov}, {Nelson},
  {Jones}, {Kern}, {Larson}, {Carey}, {Polat}, {Feng}, {Moore}, {VanderPlas},
  {Laxalde}, {Perktold}, {Cimrman}, {Henriksen}, {Quintero}, {Harris},
  {Archibald}, {Ribeiro}, {Pedregosa}, {van Mulbregt}, \& {SciPy 1. 0
  Contributors}}]{Virtanen.2020}
{Virtanen}, P., {Gommers}, R., {Oliphant}, T.~E., {et~al.} 2020, Nature
  Methods, 17, 261, \dodoi{10.1038/s41592-019-0686-2}

\bibitem[{{Wheeler} {et~al.}(2023){Wheeler}, {Valluri}, {Beraldo e Silva},
  {Dattathri}, \& {Debattista}}]{Wheeler.2023}
{Wheeler}, V., {Valluri}, M., {Beraldo e Silva}, L., {Dattathri}, S., \&
  {Debattista}, V.~P. 2023, \apj, 958, 119, \dodoi{10.3847/1538-4357/ace962}

\end{thebibliography}

%% This command is needed to show the entire author+affiliation list when
%% the collaboration and author truncation commands are used.  It has to
%% go at the end of the manuscript.
%\allauthors

%% Include this line if you are using the \added, \replaced, \deleted
%% commands to see a summary list of all changes at the end of the article.
%\listofchanges

\appendix 
\setcounter{figure}{0} % Reset figure counter
\renewcommand{\thefigure}{A\arabic{figure}} % Change

\section{The Orbit Integration Time} \label{Appendix}

To investigate the effect of integration time on our orbit classification, we repeated our analysis using an integration time of 2 Gyr, which corresponds to approximately 20 orbital periods at the end of the bar region. We employed an orbit integration with the required accuracy of $10^{-8}$. One primary concern is that a very long integration time with low integration accuracy may result in an overproduction of chaotic orbits. We examined the fraction of chaotic orbits under integration times of 20 Gyr and 2 Gyr, using the frequency drift parameter $\log _{10}(\Delta f)$ as described by \cite{Valluri2.2010}. We adopted $\log _{10}(\Delta f) > -1.2$ as the criterion for defining chaotic orbits, consistent with other studies \cite[]{Valluri.2016, Behzad.2021}.

\begin{figure*}[t]
	\centering	%
	\includegraphics[width=0.6\linewidth]{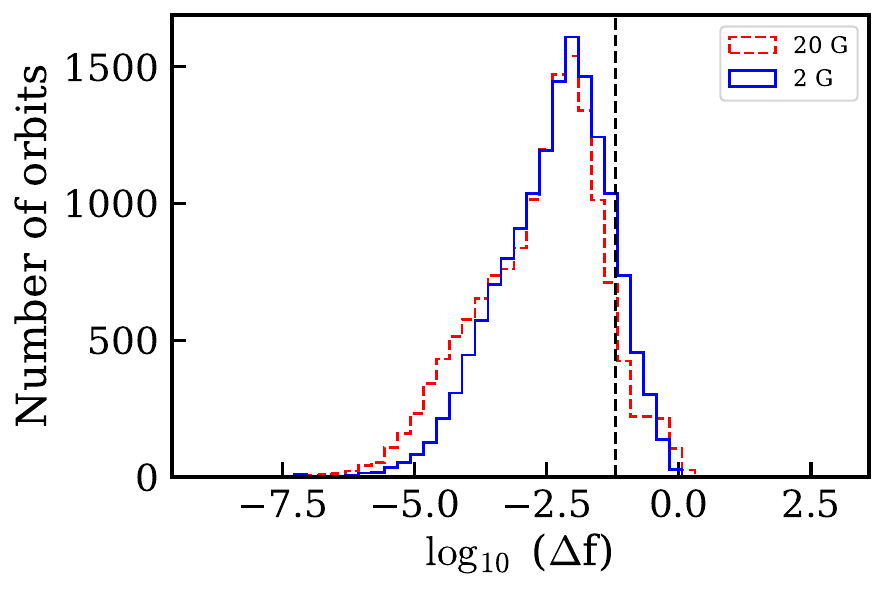}
	\caption{Distributions of the frequency drift parameter $\log _{10}(\Delta f)$ for 15,000 orbits integrated over durations of 20 Gyr (red) and 2 Gyr (blue), respectively. The dashed line indicates the threshold value of $\log _{10}(\Delta f)= -1.2$. Orbits with $\log _{10}(\Delta f)>-1.2$ are classified as chaotic.}%
	\label{fig:app1}%
\end{figure*}

Figure \ref{fig:app1} presents the distributions of the frequency drift parameter for 15,000 orbits integrated over durations of 20 Gyr (red) and 2 Gyr (blue). We found that the fraction of chaotic orbits is similar in both cases, with the orbits integrated for 2 Gyr showing approximately $2\%$ more chaotic orbits than those integrated for 20 Gyr. This difference could also be due to the randomness in selecting initial conditions, which vary for each orbit sample. This result suggests that the accuracy we are using for orbit integration is sufficient. The overall fraction of chaotic orbits in our orbit sample, as used in this paper, is $10-13\%$ across different models.

\begin{figure*}[t]
	\centering	%
	\includegraphics[width=\linewidth]{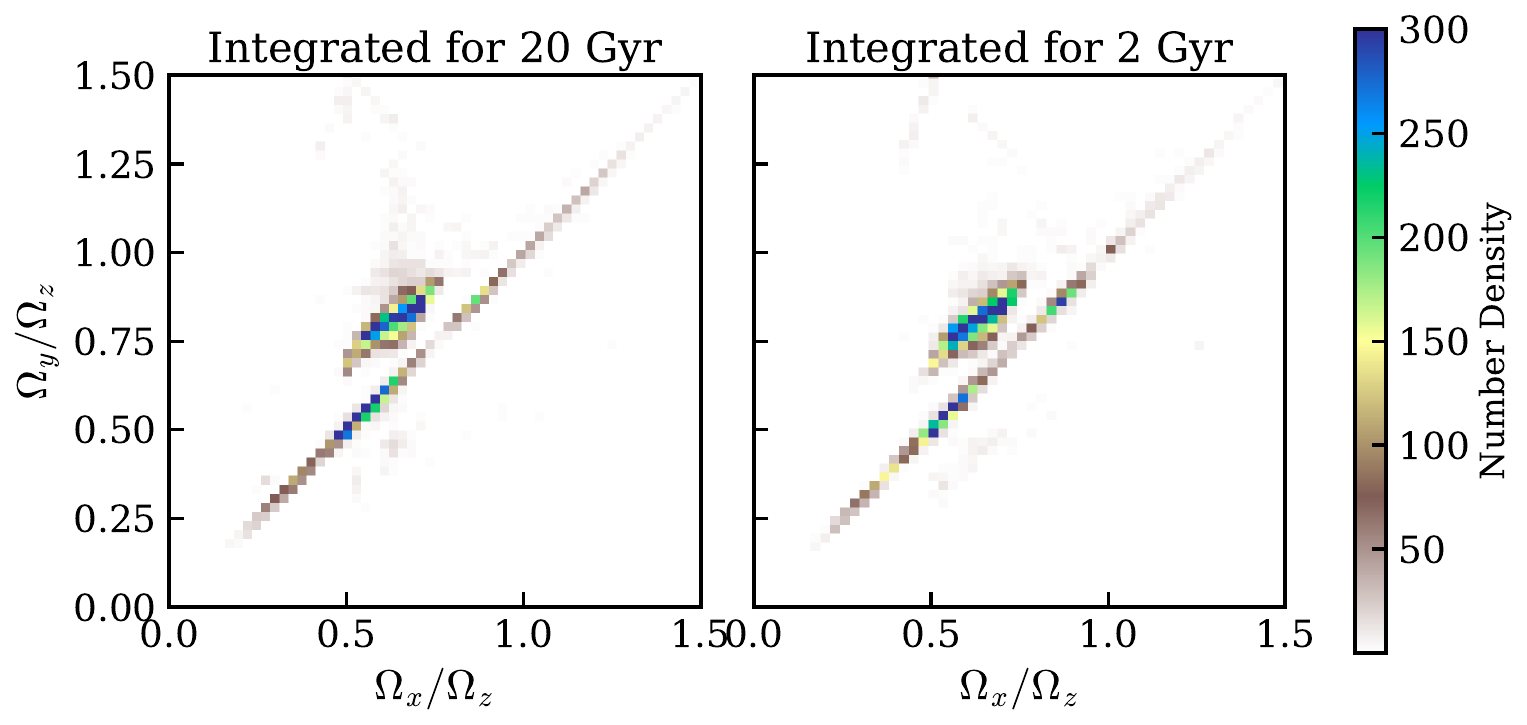}
	\caption{Comparison of our orbit frequency map for orbits integrated over durations of 20 Gyr (left) and 2 Gyr (right).}%
	\label{fig:app2}%
\end{figure*}

\begin{figure*}[t]
	\centering	%
	\includegraphics[width=0.9\linewidth]{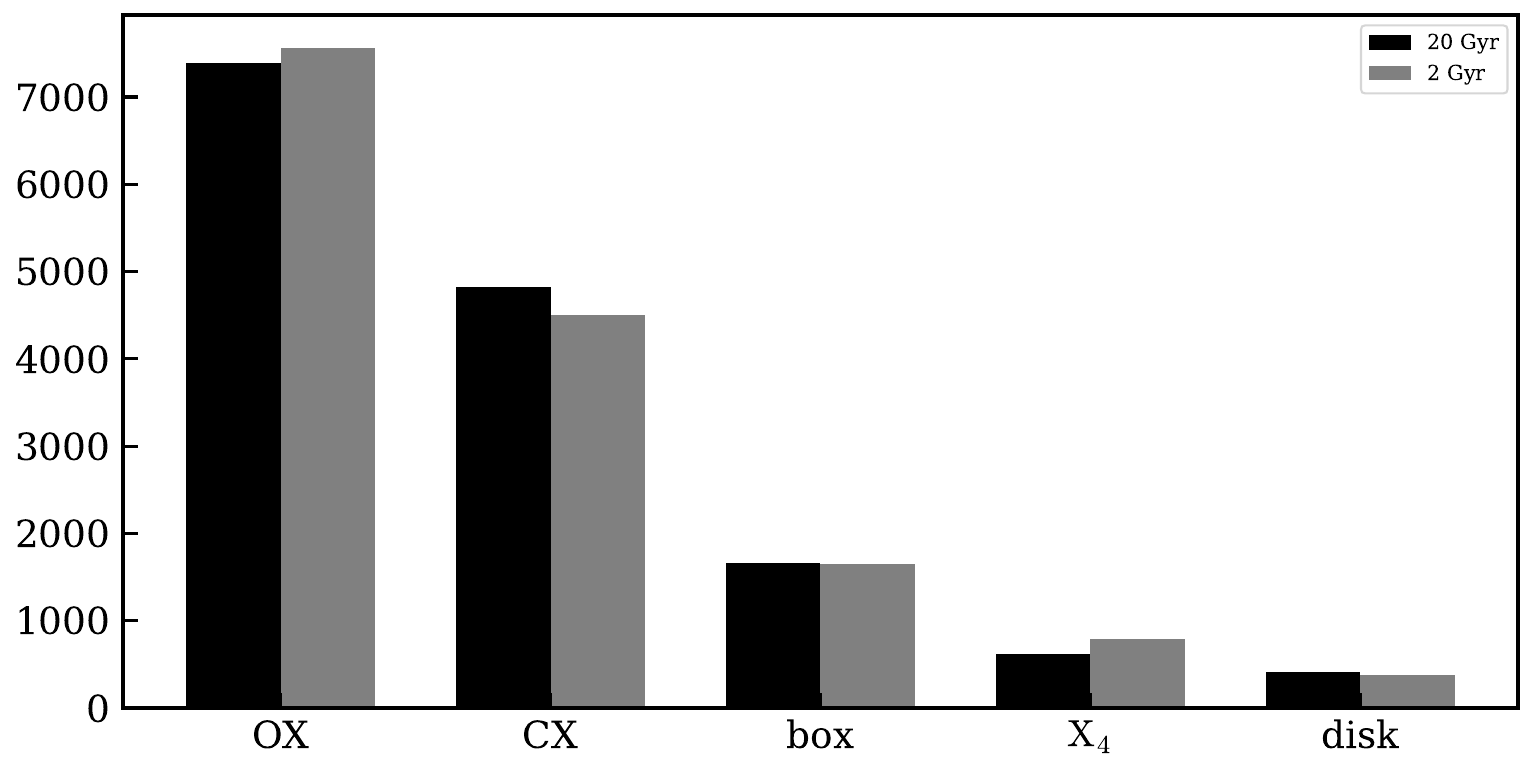}
	\caption{Comparison of our orbit classification for orbits integrated over durations of 20 Gyr (black) and 2 Gyr (gray).}%
	\label{fig:app3}%
\end{figure*}

Another concern when adopting a longer integration time is achieving more accurate orbital frequencies. However, our orbit classification is not solely based on orbit frequencies; we also consider parameters such as angular momentum, maximum z, and the ratio of maximum y to maximum x. Figure \ref{fig:app2} shows a Cartesian frequency map, colored by number density, for orbits integrated for 20 Gyr (left panel) and 2 Gyr (right panel). The frequency maps are very similar, and we did not find any systematic changes between them. Figure \ref{fig:app3} illustrates the effect of integration time on our classification, indicating only minor changes in each orbital group ($1-3\%$), which could also be due to randomness in selecting initial conditions. Overall, these tests indicate that our analysis outcomes throughout this paper are not significantly dependent on the integration time.

\clearpage

\end{document}